\documentclass[11pt]{article}
\pdfoutput=1
\usepackage{jheppub}
\usepackage{amssymb,amsmath,amstext,amsfonts}
\usepackage{bbold,ulem,bm}
\usepackage{graphics}
\usepackage{mathtools}
\usepackage{hyperref}
\usepackage{color}

\usepackage{graphicx}
\usepackage{subcaption}
\usepackage{mwe}
\usepackage{mathrsfs}
\usepackage{multirow}

\newcommand{\beq}{\begin{equation}}
\newcommand{\eeq}{\end{equation}}
\newcommand{\bal}{\begin{aligned}}
\newcommand{\eal}{\end{aligned}}
\newcommand{\bea}{\begin{eqnarray}}
\newcommand{\eea}{\end{eqnarray}}
\newcommand{\Lag}{{\mathcal{L}}}

\def\Mp{M_{{\rm Pl}}}

\def\Tdot#1{{{#1}^{\hbox{.}}}}

\def\c{c_s}

\newcommand{\F}{{\mathcal{F}}}
\def\Ez{{\cal E}_{\zeta}}

\def\eom{{\rm EoM}}
\def\d{{\rm d}}
\def\inin{\textit{in-in}\,}
\def\k{\boldsymbol{k}}
\def\a{a}
\def\b{b}
\def\h{c}
\def\g{f}
\def\A{A}

\def\chis{\chi_\star}

\begin{document}

\title{Revisiting non-Gaussianity in multifield inflation with curved field space}

\author{Sebastian Garcia-Saenz,}
\affiliation{Institut d'Astrophysique de Paris, GReCO, UMR 7095 du CNRS et de Sorbonne Universit\'e, 98bis boulevard Arago, 75014 Paris, France}
\author{Lucas Pinol}
\author{and S\'ebastien Renaux-Petel}

\emailAdd{sebastian.garcia-saenz@iap.fr}
\emailAdd{pinol@iap.fr}
\emailAdd{renaux@iap.fr}

\abstract{Recent studies of inflation with multiple scalar fields have highlighted the importance of non-canonical kinetic terms in novel types of inflationary solutions. This motivates a thorough analysis of non-Gaussianities in this context, which we revisit here by studying the primordial bispectrum in a general two-field model. Our main result is the complete cubic action for inflationary fluctuations written in comoving gauge, i.e.\ in terms of the curvature perturbation and the entropic mode. Although full expressions for the cubic action have already been derived in terms of fields fluctuations in the flat gauge, their applicability is mostly restricted to numerical evaluations. Our form of the action is instead amenable to several analytical approximations, as our calculation in terms of the directly observable quantity makes manifest the scaling of every operator in terms of the slow-roll parameters, what is essentially a generalization of Maldacena's single-field result to non-canonical two-field models. As an important application we derive the single-field effective field theory that is valid when the entropic mode is heavy and may be integrated out, underlining the observable effects that derive from a curved field space.}

\maketitle


\section{Introduction} \label{sec:intro}

Primordial non-Gaussianities are arguably the most promising probe of the physics of the early universe (see \cite{Akrami:2019izv} for the most recent observational constraints and \textit{e.g.}~\cite{Wands:2010af,Chen:2010xka,Wang:2013eqj,Renaux-Petel:2015bja,Meerburg:2019qqi} for reviews). In the context of inflation, they offer a unique observational window into energy scales even above the Hubble scale---very likely the highest scales that we may ever hope to indirectly probe. In this setting, non-Gaussianities are predominantly due to the three-point interactions of the scalar degrees of freedom that are active during the inflationary epoch, which in most typical scenarios correspond to (in a suitable gauge) the comoving curvature perturbation $\zeta$, plus possibly a slew of additional heavy fields, which are expected to be present if inflation is to be realized within a more fundamental theory, for instance as moduli arising in compactifications of string theory. If these extra fields are sufficiently heavy, with masses much larger than the Hubble scale $H$, it follows from the principle of decoupling that they should have had a negligible imprint on the inflationary dynamics, leading to the standard paradigm of single-field inflation.

However, over the past decade it has been realized that neglecting the effects of heavy fields beyond the inflaton may be premature (see \textit{e.g.}~\cite{Chen:2009we,Tolley:2009fg,Cremonini:2010ua,Achucarro:2010da,Achucarro:2012sm,McAllister:2012am,Burgess:2012dz,Arkani-Hamed:2015bza,Flauger:2016idt,Lee:2016vti,Chen:2016uwp,Chen:2018cgg,Arkani-Hamed:2018kmz}). This is best appreciated when one considers the effective field theory (EFT) of single-field inflation \cite{Creminelli:2006xe,Cheung:2007st} obtained upon integrating out all heavy fields, where one finds in particular that the remaining light degree of freedom---the curvature mode $\zeta$ in the present case---propagates with a reduced speed of sound, $c_s^2<1$. A subluminal speed of sound has direct observational consequences: for instance the tensor-to-scalar ratio is suppressed by a factor of $c_s$ relative to the fully single-field expectation. More intriguing is the effect on the bispectrum, which schematically reads $B_{\zeta}\sim \left(\frac{1}{c_s^2}-1\right){\cal O}(1)+{\cal O}(\epsilon)$, with $\epsilon$ denoting some slow-roll parameter. Considering that $\epsilon\ll1$ in slow-roll models, this result implies the striking conclusion that even a small departure from an exactly luminal dispersion relation can significantly affect the size and shape of primordial non-Gaussianities. Even more interesting however, at least from the observational viewpoint, is to entertain the possibility of a strongly reduced speed of sound, by which we mean $\frac{1}{c_s^2}-1\gtrsim1$, leading to non-Gaussianities that can be probed by next-generation experiments.

Given the expectation, purely on dimensional grounds, that $\frac{1}{c_s^2}-1=\alpha H^2/m_h^2$, for some (time-dependent) coupling parameter $\alpha$ and a typical scale $m_h$ for the heavy fields (or their perturbations to be precise), it may naively seem hard to generate a sizable bispectrum when $m_h\gg H$. However, an interesting twist in the story is provided by the coupling $\alpha$, which in two-field models (with an inflaton and a single extra heavy field) is related to the degree of geodesic deviation of the inflationary trajectory in the internal field space, as we will make explicit below. Indeed, it has recently been appreciated that having a large coupling with $\alpha\gg1$ can be easily achieved within multi-field models of inflation with a curved field space, which are characterized by non-canonical kinetic terms \cite{Renaux-Petel:2015mga,Brown:2017osf,Mizuno:2017idt,Christodoulidis:2018qdw,Garcia-Saenz:2018ifx,Bjorkmo:2019aev,Fumagalli:2019noh,Bjorkmo:2019fls,Christodoulidis:2019mkj,Christodoulidis:2019jsx,Aragam:2019khr}. In this set-up, the interplay between the curvature of the internal space and the potential of the heavy fields can give rise to novel attractor solutions featuring a large coupling $\alpha$ and hence possibly $\frac{1}{c_s^2}-1\gtrsim1$ even if $m_h\gg H$.

These as well as other recent developments in inflationary cosmology (see \textit{e.g.} \cite{Hetz:2016ics,Achucarro:2016fby,Renaux-Petel:2017dia,Achucarro:2017ing,Linde:2018hmx,Chen:2018uul,Achucarro:2018vey,Achucarro:2018ngj,Achucarro:2019pux,Grocholski:2019mot,Cicoli:2019ulk,Mizuno:2019pcm,Panagopoulos:2019ail,Bravo:2019xdo,Achucarro:2019mea}) motivate us to revisit the problem of calculating the bispectrum in two-field models of inflation with a curved field space. Our main result is the cubic action for inflationary perturbations in comoving gauge, which allows for the direct computation of correlation functions of the mode of primary observational interest, namely the curvature perturbation $\zeta$. It is worth remarking that the cubic action for general non-canonical multi-field models has already been derived in the flat gauge \cite{Elliston:2012ab}, and has been used in numerical implementations of the transport method \cite{Mulryne:2016mzv,Dias:2016rjq,Seery:2016lko,Ronayne:2017qzn,Butchers:2018hds}. Although it is in principle possible to translate these results, via a nonlinear gauge transformation, to deduce the cubic interactions for $\zeta$, in practice this is of limited use as multiple operations would still be necessary to render the sizes of the interactions manifest.
This fact is of course well known in the single-field context: given that the reduced bispectrum is proportional to $\epsilon$ (to leading order in slow-roll), one expects the coefficients of the cubic action to be suppressed by $\epsilon^2$, and yet this is far from manifest after a direct expansion or after one switches gauge. Exhibiting the ``correct'' size of the cubic vertices of $\zeta$ requires non-trivial manipulations, as first shown by Maldacena in the single-field case \cite{Maldacena:2002vr}. Our primary goal is to extend this result to a generic model with two scalars and a curved field space. We emphasize that our result is completely general: it does not use any slow-roll approximation, and it is valid for any value of the mass of the entropic (or isocurvature) perturbation. Our cubic action is therefore applicable in all types of inflationary scenarios, such as models with features, ultra slow-roll behaviours, or displaying a non-trivial evolution of fluctuations on super-Hubble scales.

An important application of our result is that the single-field EFT for $\zeta$, obtained upon integrating out the entropic mode, can be derived in a very direct way. The resulting cubic effective action provides insight into the explicit relation between the EFT of inflation and its multi-field UV completions. In particular, we have derived an explicit result for the unique Wilson coefficient that enters in the cubic action (at leading order in the derivative expansion)  and  that is not fixed by symmetry from the quadratic action, in terms of parameters defining the full two-field theory and the inflationary trajectory. Interestingly, in addition to an expected contribution from the third derivative of the potential, we find contributions to this coefficient that depend on the curvature of the internal field space, which to our knowledge had not been appreciated before.
Our result also allows one to make contact with models of k-inflation \cite{ArmendarizPicon:1999rj,Garriga:1999vw,Seery:2005wm,Chen:2006nt}, defined by a Lagrangian ${\cal L}=P(X,\phi)$ that is a generic function of a single scalar field $\phi$ and its kinetic term $X=-\frac{1}{2}(\partial\phi)^2$. Since the EFT of inflation at leading order in the derivative expansion falls in this class of models, we can use our results to (partially) reconstruct the function $P$ by relating its first few derivatives to the parameters of the two-field UV completion. Lastly, as a byproduct of our derivation of the single-field EFT, we clarify various aspects concerning the assumptions behind the validity of integrating out the entropic perturbation and the truncation of the effective action to first order in derivatives.


\section{Generalities}

In this paper, we consider general two-field non-linear sigma models of inflation, described by the action
\beq
S=\int {\rm d}^4x\sqrt{-g}\bigg[\frac{\Mp^2}{2}\,R(g)-\frac{1}{2}\,G_{IJ}(\phi)\nabla^{\mu}\phi^I\nabla_{\mu}\phi^J-V(\phi)\bigg]\,,
\label{S}
\eeq
with $G_{IJ}$ the metric of the internal field space manifold. Our convention for the Riemann tensor is
\beq
R^I_{\phantom{I}JKL}=\Gamma^I_{JL,K}+\Gamma^I_{KM}\Gamma^M_{JL}-(K\leftrightarrow L)\,,
\eeq
where we denote by $\Gamma^I_{JK}$ the corresponding Levi-Civita connection. The fact that the field space is two-dimensional allows us to write
\beq
R_{IJKL}=\frac{R_{\rm fs}}{2}\left(G_{IK}G_{JL}-G_{IL}G_{JK}\right)\,
\eeq
in terms of the field space Ricci scalar $R_{\rm fs}$.

Before considering cubic interactions in the next two sections, here we set-up our notations, describe the gauge choice and covariant parameterization of the fluctuations that we employ, and briefly review the dynamics of the background and of linear fluctuations, that will be extensively used in the rest of the paper. 


\subsection{Background}

The inflationary background is characterized by a spatially flat Friedmann-Lema\^itre-Robertson-Walker (FLRW) metric with scale factor $a(t)$ and Hubble parameter $H(t)=\dot{a}/a$, and by homogeneous scalar fields $\bar{\phi}^I(t)$. The equations of motion of the latter read ${\cal D}_t \dot{\bar{\phi}}^I  +3H  \dot{\bar{\phi}}^I +G^{IJ} V_{,J}=0$, where the time field-space covariant derivative of any field space vector $A^I$ is defined as ${\cal D}_t A^I=\dot{A}^I+\Gamma^I_{JK}\dot{\bar{\phi}}^JA^K$. As for the Einstein equations, they can be cast in the form
\beq
\dot{\sigma}^2=2\Mp^2H^2\epsilon\,,\qquad V=M_P^2H^2(3-\epsilon)\,,
\eeq
where $\epsilon \equiv -\frac{\dot{H}}{H^2}$ and we define $\dot{\sigma} \equiv \sqrt{G_{IJ}\dot{\bar{\phi}}^I\dot{\bar{\phi}}^J}$. It is useful to introduce a particular set of vielbeins along the background trajectory, the adiabatic-entropic basis defined by $e^I_{\sigma}\equiv \dot{\bar{\phi}}^I/\dot{\sigma}$ and $e^I_s$, which is orthogonal to $e^I_{\sigma}$; the ambiguity in the direction of the latter can be fixed by requiring the basis $(e^I_{\sigma},e^I_s)$ to have a definite orientation. The metric in this basis is just the identity, since by definition $G_{IJ}e^I_{\phantom{I}\hat{I}}e^J_{\phantom{J}\hat{J}}=\delta_{\hat{I}\hat{J}}$ with $\hat{I}=\sigma,s$.
The derivatives of these orthonormal vectors can be expressed as
\beq
{\cal D}_t e^I_{\sigma}=H\eta_{\perp}e^I_s\,,\qquad {\cal D}_t e^I_s=-H\eta_{\perp}e^I_{\sigma}\,,
\eeq
and we take these relations to define the bending parameter $\eta_{\perp}$ \cite{GrootNibbelink:2000vx,GrootNibbelink:2001qt}. This dimensionless parameter quantities the acceleration of the scalar fields perpendicular to their velocities, and it is a measure of the deviation of the background trajectory from a geodesic.
With these variables, the adiabatic and entropic components of the scalar eq.\ of motion simply read
\beq
\ddot{\sigma}+3H\dot{\sigma}+V_{,\sigma}=0\,,\qquad H\dot{\sigma}\eta_{\perp}+V_{,s}=0\,.
\label{background-eoms}
\eeq


\subsection{Covariant field fluctuations and gauge choice}

{\bf Covariance}.--- When going beyond the study of linear fluctuations, one should be careful to use variables that are covariant under field space redefinitions. Although not a requirement \textit{per se}, as predictions for observable quantities do not depend on particular choices of variables, it is useful and conceptually clearer to deal with covariant objects. The concern and its resolution have been first described in Ref.~\cite{Gong:2011uw} for generic multifield models. In any given gauge, the idea is to use, not the field fluctuations $\delta\phi^I=\phi^I-\bar{\phi}^I$, which do not transform covariantly, but the vector $Q^I$ living in the tangent space at $\bar{\phi}^I$ and that corresponds to the `initial velocity' of the geodesic connecting the two points labelled by $\bar{\phi}^I$ and $\phi^I$ (this geodesic is unique if the separation between the two points is sufficiently small). Up to third order in fluctuations, one finds the following relationship between the covariant perturbation $Q^I$ and the coordinate perturbation $\delta\phi^I$:\footnote{The results looks superficially different from Eq. 8 in \cite{Gong:2011uw} as they are using the covariant derivative $\Gamma^I_{JK;L}$ and not the simple derivative $\Gamma^I_{JK,L}$ as we do. The equivalent Eq.~2.4 in \cite{Elliston:2012ab} has typos and should read as ours.}
\beq
\delta\phi^I=Q^I-\frac{1}{2}\,\Gamma^I_{JK}Q^JQ^K+\frac{1}{6}(2\Gamma^I_{LM}\Gamma^M_{JK}-\Gamma^I_{JK,L})Q^JQ^KQ^L+{\cal O}(Q^4)\,.
\label{deltaphi-cubic}
\eeq 

\noindent {\bf Gauge choice and constraints}.--- The description of the mixing between the fluctuations of the scalar fields and the ones of the metric is simplified by using the ADM form of the metric \cite{adm,Salopek:1990jq}:
\beq
{\rm d}s^2=-N^2 {\rm d}t^2+h_{ij}({\rm d}x^i+N^i {\rm d} t)({\rm d}x^j+N^j {\rm d}t)\,,
\eeq
where $N$ is the lapse function and $N^i$ the shift vector, and in terms of which the action \eqref{S} reads\footnote{As we will pay attention to boundary terms, which contribute to higher-order correlation functions in general, we note that this result actually corresponds to the full action of General Relativity, composed of the Einstein-Hilbert term supplemented by the Gibbons-Hawking-York boundary term that makes the initial problem well defined, and that we omited for simplicity in Eq.~\eqref{S}.}
\beq\bal
S&=\frac{1}{2}\int {\rm d}t {\rm d}^3 x \sqrt{h}\,\left(N R^{(3)} +\frac{1}{N} (E_{ij} E^{ij}-E^2)\right) \\
&+ \frac{1}{2}\int {\rm d}t {\rm d}^3 x \sqrt{h}N\,\left(\frac{1}{N^2}G_{IJ}v^Iv^J -G_{IJ}h^{ij} \partial_i \phi^I \partial_j \phi^J-2V \right) \,,
\label{action-ADM}
\eal
\eeq
where $h=$ det$(h_{ij})$ and  $R^{(3)}$ is the Ricci curvature calculated with $h_{ij}$. The symmetric tensor $E_{ij}$ is defined by $E_{ij}=\frac12 \dot{h}_{ij}-N_{(i|j)}$, where the symbol $|$ denotes the spatial covariant derivative associated with 
the spatial metric $h_{ij}$, and $v^I=\dot \phi^I -N^j \partial_j \phi^I$.
The lapse and shift appear without time derivatives in the action and can thus be solved from their eqs.\ of motion in terms of the genuine degrees of freedom. 

Throughout the paper, we neglect tensor and vector perturbations, for the usual reasons that they are decoupled from the scalar fluctuations at linear order, and they only contribute to higher-order correlation functions of the latter by loops, which are suppressed compared to the tree-level interactions we will take into account.
For scalar fluctuations, a usual gauge choice when studying multifield inflation is the spatially flat gauge, such that $h^{{\rm flat}}_{ij}=a^2(t) \delta_{ij}$, and in which all physical degrees of freedom are the field fluctuations $Q^I$. This choice is made in many studies (see \textit{e.g.} \cite{Seery:2005gb,Langlois:2008mn,Langlois:2008qf,Tzavara:2011hn,Elliston:2012ab,Tzavara:2013wxa} for general formalisms) and has a number of advantages. However, as we discussed in the introduction, here we consider the comoving gauge in which the adiabatic field fluctuation $e_{\sigma I} Q^I$ is set to zero --- but not the entropic fluctuation\footnote{We prefer not to call this variable $Q_s$ as the latter is usually employed in the literature to refer to $e_{s} Q^I$ in the flat gauge.} $e_{s} Q^I \equiv \F$  --- and the spatial part of the metric reads $h^{{\rm comoving}}_{ij}=a^2e^{2\zeta}\delta_{ij}$. We note that referring to this gauge as comoving is a slight abuse of
terminology as there is strictly speaking no comoving gauge in multifield models beyond linear perturbation theory \cite{Rigopoulos:2005xx,Langlois:2006vv,RenauxPetel:2008gi,Lehners:2009ja}. However, as shown in these works, setting $e_{\sigma I} Q^I=0$ defines an approximate comoving gauge on super-Hubble scales in expanding universes, so we decided to keep the terminology `comoving gauge' for simplicity.

To compute the action up to cubic order, it is sufficient to plug back in the action the expressions of the lapse and the shift at linear order in terms of the physical degrees of freedom $\zeta$ and $\F$. Writing $N=1+\alpha$ and $N^i=\delta^{ij} \partial_j\theta/a^2$,
one obtains
\beq
\alpha^{(1)}=\frac{\dot{\zeta}}{H}\,,\qquad \theta^{(1)}=-\frac{\zeta}{H}+\chi\,,
\label{solution-constraints}
\eeq
where the function $\chi$ is defined by
\beq
\frac{1}{a^2}\,\partial^2\chi=\epsilon\dot{\zeta}+\frac{\dot{\sigma}\eta_{\perp}}{\Mp^2}\,\F\,.
\label{def-chi}
\eeq
Let us note that $\chi$ is not merely a useful quantity in intermediate calculations. Rather, $\chi$ defined at linear order by $\theta+\zeta/H$ in the comoving gauge that we employ coincides with $-\Psi/H$, where $\Psi$ is the gauge-invariant Bardeen potential (see \textit{e.g.} Ref.~\cite{Langlois:2008mn}). Relatedly, the solution \eqref{def-chi} to the constraint equations corresponds in our gauge with the gauge-invariant relativistic generalization of the Poisson equation $ \partial^2 \Psi/a^2=\delta \rho_{{\rm m}}/(2 \Mp^2)$, with $\delta \rho_{{\rm m}}$ the linear comoving energy density perturbation. When $\partial^2 \chi/a^2$ is negligible on super-Hubble scales, one recovers from Eq.~\eqref{def-chi} the familiar feeding of the (comoving) curvature perturbation by the entropic fluctuation $\F$ when the background trajectory differs from a geodesic ($\eta_\perp \neq 0$).


\subsection{Quadratic action and linear equations of motion}

At second order in the fluctuations, after substituting $\alpha=\alpha^{(1)}$ in the action \eqref{action-ADM} (the contributions from $\theta^{(1)}$ cancel out at this order), and using the background eq.\ of motion, ones arrives at (writing $S=\int {\rm d}t \,{\rm d}^3 x {\cal L}$)
\beq
\Lag^{(2)}=a^3\bigg[\Mp^2\epsilon\left(\dot{\zeta}^2-\frac{(\partial \zeta)^2}{a^2}\, \right)+2\dot{\sigma}\eta_{\perp}\dot{\zeta}\F+\frac{1}{2}\left(\dot{\F}^2-\frac{(\partial \F)^2}{a^2}-m_s^2\F^2\right)\bigg]\,,
\label{L2}
\eeq
where one finds the familiar expression for the entropic mass:
\beq
m_s^2\equiv V_{;ss}-H^2\eta_{\perp}^2+\epsilon H^2  \Mp^2 R_{\rm fs}\,,
\label{ms2}
\eeq
with $V_{;ss}=e_s^I e_s^J V_{;IJ}$ the projection of the covariant Hessian of the potential along the entropic direction. For future use we write the linear eqs.\ of motion deduced from $\Lag^{(2)}$:
\bea\bal
\Ez&\equiv -\frac{1}{a^3}\,\frac{\delta S^{(2)}}{\delta\zeta}= 2\Mp^2\bigg[\frac{1}{a^3}\left(a^3\epsilon\dot{\zeta}\right)^{\cdot}-\frac{\epsilon}{a^2}\,\partial^2\zeta\bigg]+\frac{2}{a^3}\left(a^3\dot{\sigma}\eta_{\perp}\F\right)^{\cdot}\, \\
{\cal E}_{\F}&\equiv -\frac{1}{a^3}\,\frac{\delta S^{(2)}}{\delta\F}=\frac{1}{a^3}\left(a^3\dot{\F}\right)^{\cdot}-\frac{1}{a^2}\,\partial^2\F+m_s^2\F-2\dot{\sigma}\eta_{\perp}\dot{\zeta}\,, \label{eoms}
\eal\eea
and we note that the eq.\ of motion of $\zeta$ can also be compactly rewritten as
\beq
\Ez=\frac{2 \Mp^2}{a^2} \partial^2\left(\dot{\chi}+H \chi-\epsilon \zeta  \right)
\label{zeta-eom-chi}
\eeq
in terms of $\chi$ defined in Eq.~\eqref{def-chi}. As the formal appearance of \eqref{zeta-eom-chi} is the same as in the single-field case, it will be particularly useful in order to extend the single-field computation of the cubic action to the two-field situation. Note eventually that the conjugate momenta of the fields read (at linear order)
\bea
p_\zeta&=&2 a^3 \Mp^2 \left(\epsilon\dot{\zeta}+\frac{\dot{\sigma}\eta_{\perp}}{\Mp^2}\,\F \right)=2 a \Mp^2 \partial^2 \chi\,, \label{pzeta}\\
p_{\F}&=&a^3 \dot{\F}\,.
\label{pF}
\eea


\section{Multifield cubic action}
\label{two-field-cubic}

Expanding the full action \eqref{action-ADM} to cubic order and substituting the linear solution \eqref{solution-constraints} for $\alpha$ and $\theta$, it is straightforward to obtain a first form of the cubic action. Details about this computation can be found in appendix \ref{expansion}. We add the three contributions \eqref{LGR3}-\eqref{Lkin3} and use the background equations to replace $\dot{\sigma}^2$, $V$, $V_{,s}$ and $V_{;ss}$ in favor of $H$, $\epsilon$, $\eta_{\perp}$ and $m_s^2$ (but single powers of $\dot{\sigma}$ are kept as is for now), finding:
\beq\bal
\Lag^{(3)}&=a^3\bigg\{\Mp^2\bigg[\epsilon\left(3\zeta-\frac{\dot{\zeta}}{H}\right)\dot{\zeta}^2-\frac{\epsilon}{a^2}\,\zeta (\partial \zeta)^2+\frac{1}{2a^4}\left(3\zeta-\frac{\dot{\zeta}}{H}\right)\left(\partial_i\partial_j\theta\partial_i\partial_j\theta-(\partial^2\theta)^2\right)\\
&\quad -\frac{2}{a^4}\,\partial_i \zeta\partial_i \theta\partial^2\theta\bigg]+\dot{\sigma}\eta_{\perp}\left(6\zeta-\frac{\dot{\zeta}}{H}\right)\dot{\zeta}\F+\frac{1}{2}\left(3\zeta-\frac{\dot{\zeta}}{H}\right)\dot{\F}^2-\frac{1}{2a^2}\left(\zeta+\frac{\dot{\zeta}}{H}\right) (\partial \F)^2\\
&\quad -\frac{1}{a^2}\,\dot{\F}\partial \F\partial \theta-\frac{1}{2H}\Big(m_s^2+2H^2\eta_{\perp}^2-2\epsilon M_P^2H^2R_{\rm fs}\Big)\dot{\zeta}\F^2-\frac{3}{2}\,m_s^2\zeta\F^2\\
&\quad -\frac{1}{6}\left(V_{;sss}-2\dot{\sigma}H\eta_{\perp}R_{\rm fs}+\epsilon \Mp^2H^2R_{{\rm fs},s}\right)\F^3\bigg\}+{\cal D}_0\,,
\eal
\label{L3-naive}
\eeq
where $\theta=-\frac{\zeta}{H}+\chi$, $V_{;sss} \equiv e_s^I e_s^J e_s^K V_{;IJK}$, $R_{{\rm fs},s}=e_s^I  R_{{\rm fs},I}$ and
\beq
{\cal D}_0=\frac{\Mp^2}{2}\,\partial_t\bigg\{2a^3\left[-9H\zeta^3+\frac{1}{a^2H}\,\zeta(\partial\zeta)^2\right]\bigg\}
\label{D0}
\eeq  
is a total derivative term. When setting $\F$ to zero, this results boils down to Eq.~(3.7) in Maldacena's classic paper \cite{Maldacena:2002vr}, as it should. However, for the same reason as there, the form \eqref{L3-naive} of the cubic action is not particularly useful, and even misleading for estimating the amplitude of non-Gaussianities. In the pure adiabatic sector for instance (the $\zeta^3$ terms in brackets), there appear cubic interactions of order $\epsilon^0$ and $\epsilon$, whereas it is known, by comparing to the cubic action computed in the flat gauge, that interactions terms in this sector are genuinely suppressed by $\epsilon^2$ (where $\epsilon$ is a generic slow-varying parameter). In the single-field case, a lot of work is required to render explicit the true size of the cubic interactions, by performing multiple integrations by parts and making use of the linear equation of motion.
Our task here is similar but more complicated as we have to deal with the two coupled fluctuations $\zeta$ and $\F$. However, by generalizing it to the two-field case, we actually simplified the computation compared to the existing literature for the single-field framework (to our knowledge, this is only presented in Ref.~\cite{Collins:2011mz}). We explain in detail the different steps of this long computation in appendix \ref{manipulations}, which is a central part of our work. Before giving our result, we would like to emphasize conceptual points related to it. 


\subsection{Principles of the computation}
\label{principles}

Let us recall generally that after having quantized the linear Gaussian theory, and identified the interacting action, one can determine higher-order correlation functions by using the \inin (also called Schwinger-Keldysh) formalism \cite{Jordan:1986ug,Calzetta:1986ey,Weinberg:2005vy}. Starting from first principles in quantum field theory, this shows that the expectation value of an observable $O(t)$ (an hermitian operator) can be computed perturbatively as
\bea
\langle in |  O(t) | in \rangle = \langle 0| \left[ \bar T \exp \left( i \int_{-\infty(1-i \epsilon)}^t H_I(t') \d t' \right) \right] O^I(t) \left[ T \exp \left( -i \int_{-\infty(1+i \epsilon)}^t H_I(t'') \d t'' \right)\right] |0\rangle  \nonumber
 \label{equ:in-in}
\eea
where we omit the hat on all operators for simplicity, $ | in \rangle$ is the vacuum of the interacting theory at some moment $t_i$ in the far past, $T$ denotes the time-ordered product, the $I$'s indicate the use of the interaction picture, the $i \epsilon$ prescription projects onto the true vacuum and turns off the interactions in the far past, and $H_I$ is the interacting Hamiltonian. At first-order in the latter, as relevant for the calculation of tree-level three-point correlation functions, one finds
\beq
\label{equ:first}
\langle O(t) \rangle^{(1)}=  i \int_{-\infty(1-i \epsilon)}^t \d t' \langle 0 | \left[ H_{I}(t'),O^I(t)\right]   | 0 \rangle   \, .
\eeq
The form \eqref{L3-naive} of the cubic action makes the use of this formalism transparent as it contains only the fields $\zeta$ and $\F$ and their first-order time derivatives. Hence, it is straightforward to determine their conjugate momenta and the interacting Hamiltonian at cubic order, which turn out to simply read 
$H^{(3)}=-\int \d^3x {\cal L}^{(3)}$, where $\dot{\zeta}$ and $\dot{\F}$ in the right-hand side should be expressed in terms of the (linear) momenta \eqref{pzeta}-\eqref{pF}. However, as explained above, we will manipulate the cubic action, performing integrations by parts that will make appear second-order time derivatives of the fields. While this is classically allowed, what is the status of these manipulations in the quantum theory?\\

There are two ways to address this question. The first is conceptually the clearest. It says that the \inin formalism should be applied to the form \eqref{L3-naive} of the cubic action and the corresponding Hamiltonian. After using the expansion of the fields into creation and annihilation operators in \eqref{equ:first} (remember that all fields here are in the interaction scheme and hence are free fields), the computation of say, the tree-level bispectrum, amounts to performing time integrals of products of mode functions, which are $c$-numbers. Doing so, one is perfectly allowed to employ integrations by parts and use that the mode functions obey $\Ez=0$ and ${\cal E}_{\F}=0$. The second practical point of view, justified by the previous reasoning, is that one can readily perform integrations by parts at the level of the classical interacting action: all second-order time derivatives generated in this process should be thought of, in the quantum theory, as operators defined in terms of the original fields and linear momenta by imposing that the linear equations of motion are satisfied. In other words, there is no independent quantum operator associated to $\ddot{\zeta}$ or $\ddot{\F}$, which would wrongly signal the existence of additional degrees of freedom beyond $\zeta$ and $\F$, and $\Ez$ and ${\cal E}_{\F}$ can be, and should be taken to be identically zero.\\
 
A last important subtlety concerns temporal boundary terms in the action. While they they do not contribute to the equations of motion, they do in general contribute to correlation functions in the \inin formalism \cite{Arroja:2011yj,Burrage:2011hd,Rigopoulos:2011eq} (contrary to computations of in-out scattering amplitudes in particle physics). From Eq.~\eqref{equ:first}, one indeed deduces that a total derivative term in ${\cal L}^{(3)}$ gives a local contribution (in time):
\beq
{\cal L}^{(3)} \supset \frac{d}{dt} B \qquad \Rightarrow \qquad  \langle O(t) \rangle^{(1)} \supset -i \int \d^3x  \langle 0 | \left[ B(t) , O(t)\right]   | 0 \rangle\,.
\label{contribution-boundary-term}
\eeq
This shows for instance that boundary operators which do not involve time derivatives do not contribute to the correlation function of an operator involving fields only (as field operators commute with themselves), like the primordial bispectrum, in agreement with a similar argument in \cite{Burrage:2011hd}. On the contrary, boundary operators involving momenta are relevant in general, and to keep the possibility to compute more general correlation functions, we will keep all temporal boundary terms. However, we will discard the spatial boundary terms generated by the numerous spatial integrations by parts, as they do not contribute to any correlation function.


\subsection{Result}
\label{result}

Referring the interested reader to the appendix \ref{manipulations} for the derivation, we quote here our final result for the cubic action:
\beq\bal \label{eq:cubic action best form1-main-text}
\Lag^{(3)}&= \Mp^2 \,a^3\left[ \epsilon (\epsilon-\eta) \dot{\zeta}^2\zeta+ \epsilon (\epsilon+\eta)\zeta \frac{\left(\partial \zeta \right)^2}{a^2}  +
\left(\frac{\epsilon}{2}-2\right)\frac{1}{a^4} \left(\partial\zeta\right)\left(\partial\chi\right) \partial^2 \chi + \frac{\epsilon}{4 a^4} \partial^2 \zeta ( \partial \chi)^2 \right] \\
&+a^3\bigg[ \frac12 m_s^2 (\epsilon+\mu_s)\zeta \F^2+\left(2\epsilon-\eta-2\lambda_{\perp}\right)\dot\sigma \eta_\perp \zeta  \dot\zeta  \F
+\frac{\dot{\sigma}\eta_{\perp}}{a^2H}\,\F(\partial\zeta)^2\\
& -\frac{\dot{\sigma}\eta_{\perp}}{H}\,\dot{\zeta}^2\F -\frac{1}{H}\left(H^2\eta_{\perp}^2-\epsilon \Mp^2H^2R_{\rm fs}\right)\dot{\zeta}\F^2-\frac{1}{6}\left(V_{;sss}-2\dot{\sigma}H\eta_{\perp}R_{\rm fs}+\epsilon \Mp^2H^2R_{{\rm fs},s}\right)\F^3\\
&+\frac12\epsilon\zeta \left(\dot{\F}^2+\frac{(\partial\F)^2}{a^2}\right)- \frac{1}{a^2}\dot{\F}\partial \F\partial \chi \bigg] +{\cal D}+{\cal E}
\eal
\eeq
with 
\beq\bal \label{D-main-text}
{\cal D}&= \frac{d}{dt}\bigg\{-\frac{a}{2 H}\zeta(\partial\F)^2+\frac{a \Mp^2}{H}(1-\epsilon)\zeta(\partial\zeta)^2-9H \Mp^2 a^3 \zeta^3 -\frac{a^3}{2H} (m_s^2+4 H^2 \eta_\perp^2)\,\zeta\F^2\\
&\quad  -\frac{\Mp^2 }{4aH^3} (\partial \zeta)^2 \partial^2 \zeta  -\frac{\zeta p_\zeta^2}{4 \epsilon H a^3 \Mp^2} +\frac{\dot{\sigma} \eta_\perp}{\epsilon H \Mp^2}  \F \zeta p_\zeta 
-\frac{\zeta}{8 a^3 H \Mp^2}\left( \partial^{-2} p_{\zeta, ij} \partial^{-2} p_{\zeta,ij}-p_\zeta^2\right) \\
& \quad+\frac{\zeta}{4aH^2}  \left(\zeta_{,ij} \partial^{-2} p_{\zeta,ij}-\partial^2\zeta p_\zeta\right)-\frac{1}{2H a^3}\,\zeta p_{\F}^2\bigg\}\,. 
\eal\eeq
and
\bea\bal
{\cal E}&=\frac{a^3}{H}\,{\cal E}_{\zeta}\bigg[\dot{\zeta}\zeta-\frac{1}{4a^2H}\left( (\partial \zeta)^2-\partial^{-2}\partial_i\partial_j(\partial_i\zeta\partial_j\zeta) -2H\left(\partial \zeta\partial \chi-\partial^{-2}\partial_i\partial_j(\partial_i\zeta\partial_j \chi)\right) \right)\bigg]+\frac{a^3}{H}\,{\cal E}_{\F}\zeta\dot{\F}\,,\,\,\,
\eal
\eea
and where it will prove convenient to use the (linear) conjugate momenta \eqref{pzeta}-\eqref{pF} in the boundary term \eqref{D-main-text}. Employing integrations by parts and making appear the linear equations of motion, the cubic action may appear in very different forms. We now argue why the result \eqref{eq:cubic action best form1-main-text} is in some sense the best form one can achieve, and emphasise a couple of points regarding it.\\

\noindent \textbullet \, First, we stress that this compact result is exact. In particular, the various parameters $\epsilon=-\dot{H}/H^2, \eta =\dot{\epsilon}/(H \epsilon), \lambda_\perp=\dot{\eta}_\perp/(H \eta_\perp), \mu_s=\Tdot{(m_s^2)}/( H m_s^2)$ are just short-hand notations, and no slow-varying approximation has been employed. It can hence be used in any model, whatever the dynamics of the background and the related mass scales dictating the physics of entropic fluctuations and their couplings to the adiabatic fluctuation. \\
\noindent \textbullet \, Contrary to the intermediate result \eqref{L3-naive}, the genuine size of interactions is made manifest in \eqref{eq:cubic action best form1-main-text}. Concerning the pure adiabatic sector, the interactions in ${\cal O}(\zeta^3)$ are explicitly of order ${\cal O}(\epsilon,\eta)^2$ or higher, like in the single-field case.
In a similar way, besides the intrinsic multifield effects encoded in operators proportional to $\eta_\perp, m_s^2, R_{\rm fs}, R_{{\rm fs},s}$ and $V_{;sss}$, the interactions in ${\cal O}(\zeta \F^2)$ in the last line are proportional to $\epsilon$.\\
\noindent \textbullet \, As we will demonstrate in the next section, this form is particularly convenient to discuss the limit of a heavy entropic fluctuation that can be integrated out, resulting in a single-field effective field theory for the adiabatic mode $\zeta$. \\
\noindent \textbullet \, How to dispatch interactions between the bulk Lagrangian and the boundary term $\cal D$ has been chosen to minimize the influence of the latter and to easily deduce their effects. In this respect, note that in the single-field case where $\F=0$ and only the first line of the bulk action is present, our result matches the form of the cubic action given in Ref.~\cite{Burrage:2011hd}. In particular, no operator in $\zeta^2 \dot{\zeta}$ is present in the bulk action, contrary to Refs.~\cite{Maldacena:2002vr,Seery:2005wm,Chen:2006nt}. As argued in \cite{Burrage:2011hd}, nothing is gained by having one additional operator in the bulk action, all the more as it would come with an extra operator in the boundary term that would contribute to the primordial bispectrum. Below, we discuss the contributions of the boundary term to the primordial bispectrum in a general multifield situation.


\subsection{Contribution of boundary terms} \label{sec:boundary terms}

It is common practice to evaluate the contributions of boundary terms by performing field redefinitions \cite{Maldacena:2002vr,Seery:2005wm,Chen:2006nt}. The idea is that by carefully choosing the field redefinition $\zeta=\tilde{\zeta}+f[\tilde{\zeta},\tilde{\F}]$ the second and cubic action written in terms of $\tilde{\zeta}$ may not contain contributing boundary terms. One can then simply evaluate the difference between the 3-point correlation functions of $\zeta$ and $\tilde{\zeta}$ by applying Wick theorem. However, as we explain in more details in appendix \ref{sec:field-redefinitions}, this procedure can be ambiguous as different field redefinitions may be chosen to cancel boundary terms, giving different answers for $\langle \zeta^3 \rangle$ in general.

In what follows, we evaluate the contribution of boundary terms following first principles. Concentrating for definiteness on the main observable of interest, the three-point correlation function of the curvature perturbation
 \beq
 \langle \zeta_{\boldsymbol{k}_1} \zeta_{\boldsymbol{k}_2} \zeta_{\boldsymbol{k}_3} \rangle \equiv (2\pi)^3 \delta(\sum_i \boldsymbol{k}_i) B_{\zeta}(k_1,k_2,k_3)\,,
 \label{Bispectrum}
 \eeq
the relevant boundary terms, that involve the conjugate momentum $p_\zeta$ and hence contribute to the bispectrum, read
\bea
S_{{\cal D}}&\supset (2 \pi)^3 \int \prod_i \frac{{\rm d}^3 \boldsymbol{k}_i}{(2 \pi)^3} \delta(\sum_i \boldsymbol{k}_i)   \bigg[  \a(\k_1,\k_2,\k_3) \F(\k_1) \zeta(\k_2)  p_\zeta(\k_3)+ \b(\k_1,\k_2,\k_3)  \zeta(\k_1) \zeta(\k_2) p_\zeta(\k_3) \nonumber \\
&+  \h(\k_1,\k_2,\k_3) p_\zeta(\k_1) p_\zeta(\k_2)  \zeta(\k_3)  \bigg]\,,
\label{SD}
\eea
where $\b$ and $\h$ are taken symmetric under $\k_1 \leftrightarrow \k_2$ without loss of generality, with 
\bea
\a(\k_1,\k_2,\k_3) &=&\frac{\dot{\sigma} \eta_\perp}{\epsilon H \Mp^2} \\
\b(\k_1,\k_2,\k_3) &=&\frac{1}{8 a^2 H^2} \left(k_1^2+k_2^2-(\k_1 \cdot \hat{\k}_3)^2-(\k_2 \cdot \hat{\k}_3)^2 \right) \\
\h(\k_1,\k_2,\k_3) &=&\frac{1}{8 a^3 \epsilon H \Mp^2} \left(-2+\epsilon \,(1-(\hat{\k}_1 \cdot \hat{\k}_2)^2) \right)\,,
\eea
and where $\hat{\k}_i$ is the unit vector $\k_i/k_i$. Using Eqs.~\eqref{equ:first}-\eqref{contribution-boundary-term} and (the quantum) Wick theorem, these various interactions contribute to the bispectrum as products of three two-point correlation functions. For instance, the second term gives
\beq
B_\zeta \supset  i  \langle \zeta(\k_1) \zeta(-\k_1) \rangle^{'} \langle \zeta(\k_2) \zeta(-\k_2) \rangle^{'}  \langle \zeta(\k_3) p_\zeta(-\k_3) \rangle^{'} \b(-\k_1,-\k_2,-\k_3) + {\rm cc} + {\rm 5\, perms.}
\eeq
where we omit hats on all operators, and use the notation $\langle A(\k_1) B(\k_2) \rangle=(2 \pi)^3 \delta(\k_1+\k_2) \langle A(\k_1) B(-\k_1) \rangle^{'}$. While 
$\langle \zeta(\k_1) \zeta(-\k_1) \rangle^{'}$ is simply the real power spectrum $P_\zeta(k_1)$, one has 
\bea
\langle \zeta(\k) p_\zeta(-\k) \rangle^{'}&=&P_{\zeta p_\zeta}(k)+\frac{i}{2}
\label{p-zeta}
\eea
where $P_{\zeta p_\zeta}(k)=\frac12 \left(\langle \zeta(\k) p_\zeta(-\k) \rangle^{'}+\langle p_\zeta(\k) \zeta(-\k) \rangle^{'} \right)$ is the real cross-spectrum, and where we used the commutation relation $\langle \zeta(\k) p_\zeta(-\k) \rangle^{'}-\langle p_\zeta(\k) \zeta(-\k) \rangle^{'} =i$. Using \eqref{p-zeta}, it is straightforward to deduce that boundary terms of the type \eqref{SD} contribute to the bispectrum as
\bea
\label{Bispectrum-boundary}
\hspace{-2em} B_{\zeta}(\k_1,\k_2,\k_3) -B_{\zeta,{\rm bulk}}(\k_1,\k_2,\k_3) 
&=& - \a(-\k_1,-\k_2,-\k_3) P_{\zeta \F}(k_1) P_{\zeta}(k_2) \,\,  +5\, {\rm perms.} \nonumber \\
&-&2\, \b(-\k_1,-\k_2,-\k_3) P_\zeta(k_1) P_\zeta(k_2) \quad\,+2\, {\rm perms.} \nonumber \\
&-&2\, \h(-\k_1,-\k_2,-\k_3) P_\zeta(k_3) P_{\zeta p_\zeta}(k_2) \,\,+5\, {\rm perms.}\,,
\eea
where $B_{\zeta, {\rm bulk}}$, generated by bulk interactions in \eqref{eq:cubic action best form1-main-text}, can be calculated using standard methods, numerically or with analytical approximations. The form \eqref{Bispectrum-boundary} has the advantage of rendering manifest that various rewritings of the boundary terms that differ by total spatial derivatives contribute in the same way to the bispectrum, as, taking into account total momentum conservation, they all define unambiguously the same functions $\b$ and $\h$ symmetric in their first two arguments. 
As we discuss in appendix \ref{sec:field-redefinitions}, this independence of the bispectrum on the precise form of the action and on spatial boundary terms is not shared by the method of field redefinitions, which is ambiguous in general.

We stress that no approximation has been used and that Eq.~\eqref{Bispectrum-boundary} can be used at any time. However, it is particularly convenient to discuss the late-time super-Hubble behaviour. In particular, it is useful not to split $P_{\zeta p_\zeta}$ into $2 a^3 \Mp^2 \left( \epsilon P_{\zeta \dot{\zeta}}+\dot{\sigma} \eta_\perp \Mp^{-2} P_{\zeta \F} \right)$. Indeed, as discussed after Eq.~\eqref{def-chi}, $p_\zeta/(2 a^3 \Mp^2)=\partial^2 \chi/a^2$ is suppressed on super-Hubble scales. In this limit, the third term in \eqref{Bispectrum-boundary} therefore gives negligible contributions, while the second one is manifestly negligible, and we are left with only the first term, that make explicitly appear $P_{\zeta \F}$. The result \eqref{Bispectrum-boundary} thus enables one to treat the single and multifield situation in a unified manner: independently of whether $\zeta$ approaches a constant or not on super-Hubble scales, only the first term contributes. In the single-field situation, it is simply absent and the boundary terms do not contribute at late times. In a multifield setup, if entropic fluctuations get exhausted by the end of inflation (and hence an adiabatic limit is reached where $\zeta$ becomes constant), the first term eventually becomes negligible, but this need not be the case in general. In these circumstances in which correlation functions have to be followed through (p)reheating, equation \eqref{Bispectrum-boundary}, with the first term in particular, would provide correct initial conditions.\\


\section{Single-field effective theories}

In sections \ref{regime-of-validity} and \ref{sec:eft-cubic-single-field}, as a first application of our general result \eqref{eq:cubic action best form1-main-text}, we discuss the particularly interesting limit of a heavy entropic fluctuation, and the single-field effective theory that results up to cubic order when the former is integrated out at leading order in derivatives. This subject has been extensively studied in the literature (see \textit{e.g.} \cite{Tolley:2009fg,Cremonini:2010ua,Achucarro:2010da,Baumann:2011su,Shiu:2011qw,Cespedes:2012hu,Achucarro:2012sm,Avgoustidis:2012yc,Achucarro:2012yr,Gwyn:2012mw,Cespedes:2013rda,Gong:2013sma,Gwyn:2014doa,Gong:2014rna,Garcia-Saenz:2018ifx,Garcia-Saenz:2018vqf,Fumagalli:2019noh}). However, it is the first time that a general formalism is derived: away from any decoupling limit, keeping all interactions and in a generic curved field space. Our result will thus unify and generalise previous partial results. In section \ref{sec:integrating-full-fields}, we consider a large class of two-field models where one may integrate out a heavy field at the level of the full action, resulting in an effective $P(X,\phi)$ Lagrangian. From this, we derive explicit expression for the parameters governing the cubic interactions of fluctuations, and show their consistency with our previous general result where heavy \textit{fluctuations} are integrated out.


\subsection{Generalities and regime of validity}
\label{regime-of-validity}

Before moving to the actual computation in the next section, here we discuss the conditions of validity and of predictivity of the EFT we will derive. Let us recall that the linear equation of motion for $\F$, ${\cal E}_{\F}=0$, reads
\begin{equation}
\left(m_s^2-\square\right)\F=2\dot{\sigma}\eta_\perp \dot{\zeta} \quad \text{  with  } \quad  \square=-\frac{\partial^2}{\partial t^2} -3H \frac{\partial }{\partial t} + \frac{\partial^2}{a^2} \,.
\end{equation}
As a first assumption that will be made more precise below, we consider situations with $m_s^2\gg H^2$,
and work out the effective action by substituting in the second and cubic order actions \eqref{L2} and \eqref{eq:cubic action best form1-main-text} the expression for $\F$ that results from solving its eq.\ of motion at leading order in the derivative expansion, that is
\beq
\F=\F_{\rm LO}\equiv \frac{2\dot{\sigma}\eta_{\perp}}{m_s^2}\,\dot{\zeta}\,.
\label{FLO}
\eeq
Independently of any derivative expansion, let us recall that for our purpose, it is sufficient to plug back in the action the solution to the \textit{linear} eq.\ of motion for $\F$, like for the lapse and shift: the quadratic correction to $\F$ coming from cubic interactions could be kept, but its total contribution to the action identically vanishes up to cubic order.

Inserting \eqref{FLO} into the quadratic action \eqref{L2}, and consistently neglecting the kinetic and gradient terms of $\F$, one readily obtains the effective quadratic Lagrangian
\beq
\Lag^{(2)}_{\rm LO}=a^3 \Mp^2 \frac{\epsilon}{c_s^2}\left(\dot{\zeta}^2- c_s^2 \frac{(\partial \zeta)^2}{a^2} \right) \,,
\label{L2-EFT}
\eeq
where $c_s^2$ such that
\beq
\frac{1}{c_s^2}-1\equiv \frac{4H^2\eta_{\perp}^2}{m_s^2}
\label{cs2}
\eeq
is the so-called speed of sound (squared) of fluctuations. Note that the linear eq.\ of motion obtained by varying \eqref{L2-EFT} consistently coincides with the  ``effective'' eq.\ of motion of $\zeta$, \textit{i.e.} the expression for ${\cal E}_{\zeta}$ \eqref{eoms} obtained after replacing $\F=\F_{\rm LO}$:
\beq\bal
{\cal E}_{\zeta, {\rm LO}}&=2 \Mp^2\bigg[\frac{1}{a^3}\left(a^3\,\frac{\epsilon}{c_s^2}\,\dot{\zeta}\right)^{\cdot}-\frac{\epsilon}{a^2}\,\partial^2\zeta\bigg]\,.
\label{effective-eom-zeta}
\eal\eeq
In a related manner, the expression of $\chi$ \eqref{def-chi} in the two-field theory  simply boils down to $\partial^2 \chi_{{\rm LO}}/a^2=\epsilon \dot\zeta/c_s^2$ in the single-field EFT. 

We note that models giving rise to $c_s^2<0$ --- an imaginary sound speed ---  have been recently studied, as well as the implications of such a non-standard framework \cite{Garcia-Saenz:2018ifx,Garcia-Saenz:2018vqf,Fumagalli:2019noh}. These models, corresponding to a large negative $m_s^2/H^2$, can be compatible with a stable background when the background trajectory strongly deviates from a geodesic. However, in these models, entropic fluctuations experience a transient tachyonic instability,  which generates large primordial non-Gaussianities \cite{Fumagalli:2019noh}. Hence, although our results in this paper formally hold in situations with $c_s^2<0$, and may be used to analyse some of these models, in the following, we implicitly and conservatively assume that $m_s^2$ and $c_s^2$ are positive.

Obviously, neglecting the gradient term $(\partial \F)^2/a^2$ compared to the mass term $m_s^2 \F^2$ in the action requires that the relevant $k$-mode enters the low-energy regime $k^2/a^2 \ll m_s^2$.
However, there also exist adiabaticity conditions bounding the time scale of variations of background quantities, owing to the fact that we also neglected the kinetic term of $\F$. Formally inverting the equation of motion for $\F$ as:
\begin{align}
\F=\left(m_s^2-\square\right)^{-1}2\dot{\sigma}\eta_\perp \dot{\zeta} &=\frac{1}{m_s^2} \sum_{i=0}^\infty \left(\frac{\square}{m_s^2}\right)^i 2\dot{\sigma}\eta_\perp \dot{\zeta}\,,
\label{F-formal}
\end{align}
this shows that the dynamics of $\F$ is reliably described by the first term in the expansion $\F_{{\rm LO}}$ \eqref{FLO} only if $\square/m_s^2 \ll 1$. In particular it requires that backgroud quantities (and the mode function of $\zeta$ that inherits any time variation of the background) evolve on time scales much larger than $m_s^{-1}$,  so as not to excite high-frequency modes that are not captured by the low-energy effective field theory \eqref{L2-EFT}. 
More precisely, let us consider for definiteness the next-to-leading order correction to $\F$ ($i=1$ in \eqref{F-formal}). Using \eqref{effective-eom-zeta}, one obtains
\begin{align}
\label{NLO}
&\square\left(\dot{\sigma}\eta_\perp \dot{\zeta} \right)=\left(1-c_s^2\right)\dot{\sigma}\eta_\perp \frac{\partial^2\dot{\zeta}}{a^2}+c_s^2\left(2+2\epsilon-2\lambda_\perp-4s\right) \dot{\sigma}H\eta_\perp \frac{\partial^2\zeta}{a^2} \nonumber \\
&-\dot{\sigma}H^2\eta_\perp \dot{\zeta}\left[\left(-3-\epsilon-\frac{\eta}{2}+\lambda_\perp+2s\right)\left(-2\epsilon-\frac{\eta}{2}+\lambda_\perp+2s\right) -\eta\epsilon-\frac{\dot \eta}{2 H}+\frac{\dot{\lambda}_\perp}{H}+2\frac{\dot s}{H}\right]
\end{align}
where $s= (c_s^2)^{\cdot}/(2Hc_s^2).$\footnote{Of course $s$ is not independent of other parameters already introduced. Explicitly,
\beq
s=(1-c_s^2)\left(\epsilon-\lambda_{\perp}+\frac{\mu_s}{2}\right)\,.
\eeq
}
Requiring that it is consistently negligible compared to the leading-order term, \textit{i.e.} $\square\left(\dot{\sigma}\eta_\perp \dot{\zeta} \right) \ll m_s^2 \dot{\sigma}\eta_\perp \dot{\zeta}$ imposes some restrictions. The first term in the right hand side of \eqref{NLO} is readily negligible, being suppressed by $(1-c_s^2) k^2/(a^2m_s^2)$. However, the third term is safely negligible only if $H^2/m_s^2$ multiplied by the expression between brackets in \eqref{NLO} is much smaller than unity. Barring cancellations, this requires
\beq
\left(\frac{\dot{X}}{m_s X}\right)^2 \ll1 \quad \textrm{and} \quad \frac{\ddot{X}}{m_s^2 X}\ll 1
\eeq 
for the various parameters $X=(H,\epsilon,c_s,\eta_\perp)$. As a prolonged phase of inflation requires $\epsilon$ and $\eta \ll 1$, these conditions are usually taken for granted for the first two parameters, but should be kept in mind in case of transient features. More interesting is the fact that the speed of sound $c_s$ and the bending parameter $\eta_\perp$ are allowed to vary significantly on a Hubble time scale, although not on scales $m_s^{-1}$ \cite{Cespedes:2012hu}. Note that considering higher-order terms in the expansion \eqref{F-formal} would have bounded any $X^{(n)}/(m_s^n X)$, so that it is really the time scale of variation of the various quantities that are bounded, and not specific time-derivatives. Eventually, as $c_s^2$ \eqref{cs2} not only depends on the bending parameter but also on the entropic mass \eqref{ms2}, which may vary on different time scales, the validity of the EFT can not be judged simply by inspecting the time variations of $c_s$, but should be verified at the level of the various entropic quantities.
While we considered so far the first and last terms in \eqref{NLO}, the second one in $\partial^2 \zeta$ can not be straightforwardly compared to $\dot{\zeta}$ without some knowledge about the time-dependence of the mode function, which is not known analytically in a general time-dependent background. 
However, when the adiabaticity conditions above are satisfied, one expects the standard behaviours $\dot{\zeta}\sim \frac{kc_s}{a}\zeta$ inside the sound horizon and $\dot{\zeta}\sim \frac{k^2c_s^2}{a^2H}\zeta$ outside. This enables one to deduce that the second term is also well negligible under these circumstances.

Besides the adiabaticity conditions, we have seen that higher-derivative corrections to the leading-order EFT are in $k^2/(a^2 m_s^2)$. To express this in terms of energy scales $\omega$, let us use the dispersion relation $\omega^2=c_s^2 k^2/a^2$ deduced from the EFT \eqref{L2-EFT}. This shows that corrections to the EFT action are in $\omega^2/\omega_{{\rm new}}^2$, where
\beq
\omega_{{\rm new}}^2=m_s^2 c_s^2
\eeq
has been indeed identified as the energy scale of new physics above which higher-order derivative terms should be taken into account \cite{Cremonini:2010ua,Baumann:2011su}.

In retrospect, using the expression \eqref{FLO}, one can compute $\dot{\F}$:
\beq
\dot{\F}_{\rm LO}=\frac{2\dot{\sigma}\eta_{\perp}}{m_s^2}\bigg[\frac{c_s^2}{a^2}\,\partial^2\zeta-H \dot{\zeta}\left(3+\epsilon+\frac{\eta}{2}-\lambda_{\perp}+\mu_s-2s\right)\bigg]\,,
\label{Fdot}
\eeq
omitting ${\cal E}_{\zeta, {\rm LO}}$ on the right-hand side, and deduce that under the conditions of validity of the EFT, we were indeed allowed to neglect the kinetic term of $\F$ in the action compared to the mass term, the former being suppressed by the hierarchies $H^2/m_s^2$, $c_s^2 \omega^2/\omega_{{\rm new}}^2$ and combinations thereof. This shows however that any appearance of $\dot{\F}$ in the higher-order action may not be neglected, in particular if it is multiplied by possibly large factors like $m_s^2/H^2$ or $\eta_\perp$. We have taken this into account to choose the `best form' of the cubic action we displayed in \eqref{eq:cubic action best form1-main-text} (see appendix \ref{manipulations-F}).\\

Eventually, let us consider a situation in which $H^2/c_s^2\gg m_s^2$. Then modes described by the EFT, such that $k^2/a^2 \ll m_s^2$, 
are outside the sound horizon and already constant, hence the EFT can not predict the final observable values of their correlation functions. To be able to impose proper initial conditions from the EFT alone (for instance and typically Bunch-Davies), without knowledge from its two-field UV completion, there should exist an intermediate regime in which modes are both of sufficiently low-energy that they can be described by the EFT, and sufficiently under the sound horizon, \textit{i.e.} they should verify $m_s^2 \gg k^2/a^2 \gg H^2/c_s^2$. Hence, one deduces that in addition to the conditions of validity that we delineated above, one should require
\beq
\frac{H^2}{m_s^2 c_s^2} \ll 1
\label{predictivity}
\eeq
for the EFT to have predictive power, which is more constraining than simply having a massive field: $H^2/m_s^2 \ll 1$. This is of course consistent with higher-order derivative corrections to the EFT action being in $\omega^2/\omega_{{\rm new}}^2$: as modes become constant around $\omega \sim H$, describing this process with our low-energy EFT requires $H^2 \ll \omega_{{\rm new}}^2$, \textit{i.e.} Eq.~\eqref{predictivity}.


\subsection{Single-field effective theory of fluctuations}
\label{sec:eft-cubic-single-field}

To deduce the effective cubic action, we should substitute $\F$ by $\F_{\rm LO}$ \eqref{FLO} in the cubic action. As we mentioned, the form \eqref{eq:cubic action best form1-main-text} is particularly convenient for that purpose. Indeed, the last line of the bulk action should be consistently discarded, as it involves either gradient terms suppressed by $k^2/(a^2 m_s^2)$, or interactions in $\dot{\F}$ that are suppressed by $H^2/m_s^2$. 
The other bulk interactions involve $\F$ only, and their contributions to the effective action for $\zeta$ are straightforward to deduce, with $\partial^2 \chi/a^2 \to \epsilon \dot\zeta/c_s^2$ for the first line, the first two terms of the second line giving $\dot{\zeta}^2 \zeta$ interactions, the last term of the second line giving a vertex in $\dot{\zeta} (\partial \zeta)^2$, and the whole third line giving rise to interactions in $\dot{\zeta}^3$. 
As for the boundary terms, the general expression \eqref{Bispectrum-boundary} in the multifield situation shows that their contributions on super-Hubble scales are negligible in the single-field effective theory where $\zeta$ becomes constant on super-Hubble scales. Hence quoting the bulk action only for simplicity, one arrives at
\beq\bal \label{eft-cubic}
\Lag^{(3)}_{{\rm EFT, bulk}}&= \Mp^2 \,a^3 \frac{\epsilon}{c_s^2}\left[\g_0 c_s^2 \frac{\dot{\zeta}}{H} \frac{\left(\partial\zeta\right)^2}{a^2}+\frac{\g_1}{H} \dot{\zeta}^3 + \g_2 \dot{\zeta}^2\zeta + \g_3 c_s^2 \zeta \frac{\left(\partial\zeta\right)^2}{a^2}  + \g_4 \dot{\zeta} \partial_i\partial^{-2}\dot{\zeta}\partial_i\zeta + \g_5 \partial^2\zeta (\partial_i \partial^{-2} \dot{\zeta})^2\right] 
\eal
\eeq
with
\bea
\g_0&=&\left(\frac{1}{c_s^2}-1\right)  \qquad \g_1=\left(\frac{1}{c_s^2}-1\right) \A \qquad \g_2=\epsilon-\eta+2s \nonumber \\
   \g_3&=&\epsilon+\eta  \qquad \qquad \,  \g_4=\frac{\epsilon}{2 c_s^2}(\epsilon-4)  \qquad  \,\,\,\,\,\g_5=\frac{\epsilon^2}{4c_s^2} 
\label{fi}
\eea
and
\beq
\A=-\frac12(1+\c^2)+\frac23(1+2 \c^2)\frac{\epsilon H^2 \Mp^2 R_{{\rm fs}}}{m_s^2}-\frac16 (1-\c^2)\left( \frac{\kappa V_{;sss} }{m_s^2} + \frac{\kappa \epsilon H^2 \Mp^2 R_{{\rm fs},s} }{m_s^2}\right)\,,
\label{A}
\eeq
and where we introduced the so-called bending radius of the trajectory $\kappa=\sqrt{2 \epsilon} \Mp/\eta_{\perp}$. The cubic action \eqref{eft-cubic}-\eqref{A} constitutes the second main result of this paper. It incorporates without approximations all the interactions in the effective field theory that results from integrating out heavy entropic fluctuations at leading-order in a derivative expansion. The normalisation of operators in \eqref{eft-cubic} have been chosen such that the coupling constants $\g_i$'s in \eqref{fi} represent the genuine typical contributions of each interaction to the dimensionless shape function of the bispectrum, \textit{i.e.} the $\g_i$'s represent the typical contributions to $f_{{\rm NL}}$ in featureless models.
As we explain below, the set of six operators in \eqref{eft-cubic} are redundant. However, this form of the cubic action has the appealing
physical advantage to make transparent the link with the single-field effective field theory of fluctuations \cite{Creminelli:2006xe,Cheung:2007st}.\\

\noindent {\bf Dominant interactions and EFT of inflationary fluctuations}.--- As we have discussed in section \ref{regime-of-validity}, the EFT is perfectly valid in situations with $(\epsilon,\eta,s)={\cal O}(1)$, as motivated by transient features and sharp turns in field space in the multifield theory, and in which case all interactions in \eqref{eft-cubic} should be kept a priori. Here however, we concentrate on smooth models in which a slow-varying approximation $(\epsilon,\eta,s) \ll1$ is verified. In this context, a particularly interesting regime, theoretically and observationally, corresponds to situations in which the sound speed \eqref{cs2} differs substantially from unity, \textit{i.e.} $\left(\frac{1}{c_s^2}-1\right) \gtrsim {\cal O}(1)$. Considering $\A$ of order one (we will come back to this below and discuss the various contributions to $\A$), one deduces from \eqref{fi} that the first two operators in \eqref{eft-cubic} are dominant. 
Hence, not writing explicitly boundary terms, which do not play a role for observational predictions in this context, one can effectively write
\begin{equation}
S^{(3)}_{{\rm EFT, main}}=\int \d t\, \d^3 x a^3 \Mp^2 \frac{\epsilon}{H} \left(\frac{1}{c_s^2}-1\right)\left[ \dot{\zeta} \frac{\left(\partial\zeta\right)^2}{a^2}+\frac{A}{c_s^2} \dot{\zeta}^3 \right]\,.
\label{S3-EFT-main}
\end{equation}
These two operators are precisely the ones encountered in the decoupling limit of the effective field theory of single-clock inflationary fluctuations at leading-order in derivatives \cite{Cheung:2007st}: $\dot{\zeta} \left(\partial\zeta\right)^2$ interactions, whose size is fixed by symmetry in terms of the sound speed appearing in the second-order action \eqref{L2-EFT}, and $\dot{\zeta^3}$ interactions, whose overall contribution is not fixed. 
To be more precise, although the EFT of single-clock inflationary fluctuations can be in principle formulated exactly in terms of the comoving curvature perturbation $\zeta$, in practice computations in this context are often performed in terms of the Goldstone boson of spontaneously broken time translations $\pi$, in the decoupling limit in which its mixing with gravity can be neglected, at leading-order in a slow-varying approximation, and with the linear approximate relation $\zeta \simeq -H \pi$. Under these hypotheses and approximations, it agrees with Eq.~\eqref{S3-EFT-main}. However, we stress that our general result \eqref{eft-cubic}, of which \eqref{S3-EFT-main} is a particular limit, holds without these approximations: it encompasses the usual formulation of the single-clock EFT of inflation, but goes beyond it. 
To our knowledge, this is the first time that the dominant operators expected in the EFT of single-clock fluctuations are derived from first principles from a UV completion (here two-field models with a background trajectory deviating from a geodesic motion), away from the decoupling limit, and readily in terms of the observable comoving curvature perturbation. 

Under the slow-varying approximation used above, the observational predictions corresponding to the EFT \eqref{L2-EFT}-\eqref{S3-EFT-main} are very well known, for the power spectrum and bispectrum. We reproduce these results for completeness. The primordial power spectrum reads
\beq
\frac{k^3}{2 \pi^2}P_\zeta(k)= \left( \frac{H^2}{8 \pi^2 \epsilon c_s} \right)_\star\,,
\label{power-spectrum-cs}
\eeq
with a mild scale dependence given by the slight dependence of $H, \epsilon$ and $c_s$ on the time of evaluation $\star$ such that $k c_s=a H$, \textit{i.e.} $n_s-1=-(2 \epsilon+\eta+s)_\star$. As for the bispectrum \eqref{Bispectrum}, its shape $S$, such that 
\beq
B_\zeta = (2\pi)^4  \frac{S(k_1,k_2,k_3)}{(k_1 k_2 k_3)^2} A_s^2\,,
\eeq
with $A_s$ the power spectrum \eqref{power-spectrum-cs} evaluated at a pivot scale, reads, in a scale-invariant approximation: 
\bea
\hspace{-2em} S&=&S_{\dot{\zeta} \left(\partial\zeta\right)^2}+S_{\dot{\zeta}^3}\,, \qquad {\rm with} \label{S} \\ 
\hspace{-2em}S_{\dot{\zeta} \left(\partial\zeta\right)^2}&=&\left(\frac{1}{c_s^2}-1\right)  \frac{1}{k_1k_2k_3}
\left[-\frac{1}{K}\sum_{i>j}k_i^2k_j^2+ \frac{1}{2K^2}
\sum_{i\neq j}k_i^2k_j^3+\frac{3}{2K^3} \prod_i k_i^2+\frac{1}{8}\sum_{i}k_i^3 \right]
\label{Ac} \\
\hspace{-2em}S_{\dot{\zeta}^3}&=&\A \left(\frac{1}{c_s^2}-1\right) 
\frac{3k_1k_2k_3}{2K^3}\,,
\label{Alam}
\eea
and were $K=k_1+k_2+k_3$. These two individual shapes are similar, peak on equilateral triangles, and can be represented in a first approximation by the well known equilateral template \cite{Creminelli:2005hu}. However, their linear combination assumes a very different shape in the range $3.1 \lesssim \A \lesssim 4.3$, peaking near flattened configurations $k_2+k_3 \simeq k_1$, and is more accurately described by the orthogonal template \cite{Senatore:2009gt}. In this respect, it is instructive to quote the amplitude of the total shape function \eqref{S} in the representative equilateral and squashed configurations
\bea
S(1,1,1)&=& -\frac{17}{72}\left(\frac{1}{c_s^2}-1\right) \left(1-\frac{4 \A}{17}\right)  \label{S111}\\
S(1,1/2,1/2)&=&-\frac{3}{64}\left(\frac{1}{c_s^2}-1\right) \left(1-\A\right) \label{Ssquashed} \,,
\eea
where \eqref{S111} indicates in a simple manner that for values of $\A$ around $17/4=4.25$, one can not expect indeed the equilateral template to faithfully represent the bispectrum. \\

\noindent  {\bf Full result and comparison with P(X) models}.--- We derived the low energy EFT \eqref{L2-EFT}-\eqref{eft-cubic} at lowest order in derivatives in terms of the single fluctuating degree of freedom $\zeta$. Hence, it should correspond to a particular case of the EFT of single-clock inflation at lowest-order in derivatives, and indeed we have seen above that our result encompasses the decoupling limit result of this formulation. There exist another interesting class of models, single-field inflationary models of Lagrangian $P(X=-\frac12 (\partial \phi)^2,\phi)$, also-called $k$-inflation or $P(X)$ theories, whose fluctuations can be exactly described by the single-clock EFT at lowest order in derivatives (this is readily visible in the uniform inflaton gauge). Hence, our full result \eqref{L2-EFT}-\eqref{eft-cubic} should agree with the well known full quadratic and cubic action of $k$-inflation \cite{Seery:2005wm,Chen:2006nt}, upon identification of the multifield background quantities $c_s^2$ and $\g_i$'s \eqref{fi} with suitable combination of the derivatives $P^{(n)}(X)$. 

As discussed in \ref{result}, given the way we organised the splitting between bulk and boundary terms in \eqref{eq:cubic action best form1-main-text} and in the resulting EFT, \textit{i.e.} without a bulk operator in $\zeta^2 \dot{\zeta}$, it is convenient to compare our result to the one of Ref.~\cite{Burrage:2011hd}, where the same choice is made and boundary terms are innocuous. Concentrating on bulk terms, their result contains the five last operators in \eqref{eft-cubic}, but with different coupling constants, and without the first operator in $\dot{\zeta} \left(\partial\zeta\right)^2$. However, the latter can be manipulated as (see \textit{e.g.} \cite{RenauxPetel:2011sb})
\beq\bal
&\frac{a\epsilon}{H}\left(\frac{1}{c_s^2}-1\right)\dot{\zeta}(\partial\zeta)^2=\frac{a\epsilon}{c_s^2}\left((1-c_s^2)(1+\epsilon+\eta)-2s\right)\zeta(\partial\zeta)^2+\frac{a^3\epsilon(1-c_s^2)}{c_s^4}\,\frac{\dot{\zeta}^3}{H}\\
&\quad +\frac{a^3\epsilon}{c_s^4}\left((1-c_s^2)(-3+\epsilon-\eta)-2c_s^2s\right)\dot{\zeta}^2\zeta-\partial_t\bigg[\frac{a\epsilon}{H}\left(\frac{1}{c_s^2}-1\right)\zeta(\partial\zeta)^2+\frac{a^3\epsilon(1-c_s^2)}{Hc_s^4}\,\dot{\zeta}^2\zeta\bigg]\\
&\quad +\frac{a^3(1-c_s^2)}{\Mp^2Hc_s^2}\,\dot{\zeta}\zeta{\cal E}_{\zeta,{\rm LO}}\,,
\eal\eeq
where the generated boundary terms are also innocuous on large scales. Using this redundancy between operators, the coupling constants in the bulk action may be reshuffled as
\bea
\tilde{\g}_0&=&0 \qquad \qquad \tilde{\g}_1=\left(\frac{1}{c_s^2}-1\right) (1+A) \qquad  \quad\tilde{\g}_2=-3 \left( \frac{1}{c_s^2}-1 \right)+\frac{\epsilon-\eta}{c_s^2} \nonumber \\
   \tilde{\g}_3&=&\frac{1}{c_s^2}-1+\frac{\epsilon+\eta-2s}{c_s^2}  \qquad  \,  \tilde{\g}_4=\frac{\epsilon}{2 c_s^2}(\epsilon-4)  \qquad  \,\,\,\,\,\tilde{\g}_5=\frac{\epsilon^2}{4c_s^2} \,,
\label{k-inflation-fi}
\eea
which indeed exactly matches the k-inflationary result of Ref.~\cite{Burrage:2011hd} (Eqs.~3.10-3.11 there with slightly different notations), upon the identifications:
\bea
\left(\frac{1}{c_s^2}-1\right)^{{\rm P(X)}}= \frac{2X P_{,XX}}{P_{,X}} \quad &\leftrightarrow& \quad \left(\frac{1}{c_s^2}-1\right)^{{\rm two-field}}=\frac{4H^2\eta_{\perp}^2}{m_s^2} \label{cs-cs}\\
2 \frac{\lambda}{\Sigma} =\frac{2 X^2 P_{,XX}+4/3 X^3 P_{,XXX}}{X P_{,X}+2X^2 P_{,XX}} \quad  &\leftrightarrow& \quad -\left(\frac{1}{\c^2}-1 \right)A\,.\label{P(X)-A}
\eea
While \eqref{cs-cs} is obvious from the identifications of the sound speed in the two different class of models already at the level of the quadratic action, Eq.~\eqref{P(X)-A} is new and non-trivial, and we will perform an additional consistency check of it in \ref{sec:integrating-full-fields}.

Eventually, we verified that not only the bulk terms but the whole cubic actions agree between our EFT result and the $k$-inflationary one in Ref.~\cite{Burrage:2011hd}, as it should. Also, note that one can readily integrate out the heavy fluctuation $\F$ at the level of Eq.~\eqref{L3-naive}. The subsequent manipulations required to display the genuine size of interactions are similar to the ones performed in appendix \ref{manipulations} and we have checked that the result obtained in this way agrees with our computation.\\

{\bf Ultraviolet sensitivity and observable effects of curved field space}.--- As we have just seen, at the level of the effective action for the fluctuations, beyond the well known appearance of a reduced sound speed, related to the deviation of a background trajectory from a geodesic \eqref{cs2}, the precise multifield origin of the EFT is encapsulated in the dimensionless coefficient $\A$ \eqref{A}, which enters into the cubic action and hence in the non-Gaussian signal, together with $c_s^2$ (see \textit{i.e.} \eqref{S}-\eqref{Ssquashed}). The first contribution to $A$, fixed by $c_s^2$, has already been identified in a decoupling limit analysis in Ref.~\cite{Achucarro:2012sm}. The three other contributions have not been taken into account so far in a generic context. The third term, proportional to $V_{;sss}$, agrees with the decoupling limit analysis of a specific two-field model in flat field space in \cite{Gong:2013sma}. The two others, proportional to the Ricci curvature of the two-dimensional field space, as well as to its derivative along the entropic direction, are specific to models with curved field space and are newly derived. Although these geometrical quantities affect observables only through the global combination $A$, and hence their effects may be hard to disentangle from other effects like the one of the potential, we find it very interesting that the non-Gaussian signal carries such information about the field space geometry. In this respect, note that all terms in \eqref{A} are important in general, despite the fact that we integrated out heavy fluctuations and that some may naively appear suppressed by $1/m_s^2$: just like $c_s^2$ in \eqref{cs2} may differ significantly from unity when $H^2 \eta_\perp^2 \gg m_s^2$, the last three contributions involve ratios between $m_s^2$ and other physical scales than $H^2$, and hence can contribute to $\A$ as importantly as, or larger than, the first term of order one. This is clearly visible for the second contribution for instance, whose size is set by the relative contribution of the geometrical term to the entropic mass \eqref{ms2}. Eventually, we pointed out in section \ref{regime-of-validity} that, within the regime of predictivity of the EFT, relative corrections to its predictions are of order ${\cal O}(H^2/(m_s^2 c_s^2))$. Hence, contributions to $A$ of that order should be self consistently neglected.


\subsection{When full fields can be integrated out}
\label{sec:integrating-full-fields}

In some two-field models of the type we consider, one may be able to integrate out a \textit{heavy field} at the level of the full action, resulting at lowest order in derivatives in an effective EFT for a single scalar field which is of $P(X,\phi)$ type \cite{Tolley:2009fg}. In their common domain of validity, the action governing the fluctuations in these theories should agree with our general result \eqref{k-inflation-fi}-\eqref{P(X)-A} where the background is studied at the two-field level and \textit{heavy fluctuations} about it are integrated out. We perform this consistency check in a large class of models, deriving useful explicit results for observables readily in terms of the functions defining the two-field Lagrangian.

We consider the general class of Lagrangians
\beq
{\cal L}=-\frac12 e^{2b(\chi)} (\partial \phi)^2 -\frac{1}{2}(\partial \chi)^2-V(\phi,\chi) \,,
\label{general-class}
\eeq
which has been used in the past by many authors to study the effects of non-trivial kinetic terms and field space curvature (see, {\it e.g.}\ \cite{Starobinsky:2001xq,DiMarco:2002eb,DiMarco:2005nq,Lalak:2007vi,Tolley:2009fg,Pinol:2018euk}). Following Ref.~\cite{Tolley:2009fg}, one consider situations in which the effective mass of $\chi$ (called the gelaton in this reference) is much larger than the Hubble scale $H$, so that it adiabatically follows the minimum of its effective potential, at the value $\chis(\phi,X)$ that depends on the inflaton field $\phi$ and its kinetic energy $X=-\frac12 (\partial \phi)^2$. With the equation of motion of $\chi$ reading $\square \chi-2 b' e^{2b} X+V_{,\chi}=0$, one deduces that $\chis(\phi,X)$ verifies
\beq
V_{,\chi}(\phi,\chis)-2 b'(\chis) e^{2b(\chis)} X=0\,.
\label{effective-minimum}
\eeq
Substituting this back into the action, and consistently neglecting the kinetic term of $\chi$, one obtains 
\beq
{\cal L}^{{\rm EFT}}= e^{2b(\chis(\phi,X))} X -V(\phi,\chis(\phi,X))=\frac{V_{,\chi}(\phi,\chis(\phi,X))}{2 b'(\chis(\phi,X))}-V(\phi,\chis(\phi,X)) \,,
\label{L-k-inflation}
\eeq
and hence a low-energy effective theory which is equivalent, at leading order, to a $P(X,\phi)$ theory. Like in any inflationary setup, model-dependent quantum corrections to this classical picture may be important in general. Additionally, the requirement that the mass of the gelaton be both much larger than $H$ and smaller than the cutoff of the $P(X)$ theory, so that perturbation theory remains weakly coupled, imposes constraints on the parameter space of viable and observationally interesting models \cite{Tolley:2009fg,Butchers:2018hds}. Here however, we only want to check the formal consistency between predictions derived from \eqref{L-k-inflation} and the ones from our two-field reasoning, hence we keep $b(\chi)$ and $V(\phi,\chi)$ general.

In order to determine the explicit expressions of the various multifield quantities used in this paper in terms of $b, V$ and their derivatives, one only needs to know the background velocities of fields. From \eqref{effective-minimum}, one finds $V_{,\chi}=b' e^{2b} \dot{\phi}^2$ while one should consider $\dot{\chi} \simeq 0$ for consistency. Hence, in the coordinate basis $(\phi,\chi)$, the adiabatic and entropic vectors simply read
\beq
e_\sigma^I=(e^{-b},0)\,, \quad  e_s^I=(0,-1)\,, 
\label{basis-vectors}
\eeq
where we took $\dot \phi>0$ without loss of generality. Using $V_{;ss}=V_{,\chi \chi}, H^2 \eta_\perp^2=V_{,\chi} b', R_{\rm fs}=-2(b'^{2}+b''), V_{;sss}=-V_{,\chi \chi \chi}$, one obtains
\bea
m_s^2=V''-\frac{V'}{b'} (2b'^{2}+b'')
\eea
and the expressions \eqref{cs-gelaton}-\eqref{A-P(X)} below for $c_s^2$ and $A$, where here in and in the following, all derivatives of the potential are with respect to $\chi$ so that there is no source of confusion.\\

In the $P(X)$ perspective, the explicit expression of $\chis(\phi,X)$ solution of \eqref{effective-minimum} is not known in general, and hence neither is the expression of $P(X,\phi)$ in \eqref{L-k-inflation}. However, taking the derivative of \eqref{effective-minimum} with respect to $X$, one obtains
\beq
\frac{\partial \chis(\phi,X)}{ \partial X}=\frac{2 e^{2 b} b'}{V''-\frac{V'}{b'} (2b'^{2}+b'')}\,,
\eeq
which can be used to compute from \eqref{L-k-inflation} all successive derivatives $\partial P^{(n)}/ \partial X^n$, and hence the expressions of $c_s^2$ and $\lambda/\Sigma$ as defined in the left-hand sides of \eqref{cs-cs}-\eqref{P(X)-A}. Working this out, one obtains
\beq
\left(\frac{1}{c_s^2}-1\right)^{{\rm P(X)}}=\left(\frac{1}{c_s^2}-1\right)^{{\rm two-field}}=\frac{4 b' V'}{V''-\frac{V'}{b'} (2b'^{2}+b'')}
\label{cs-gelaton}
\eeq
and
\beq
2 \frac{\lambda}{\Sigma}= -\left(\frac{1}{\c^2}-1 \right)A=-\frac{2 b'^2 V'
   \left(3 b''^2 V'^2+4b'^4V'^2+b'^2 \left(2 V''' V'-3V''^2\right)-2
   b''' b' V'^2\right)}{3
   \left(2 b'^2 V'+b'' V'-b' V''\right)^2
   \left(2b'^2 V'-b'' V'+b' V''\right)}
   \label{A-P(X)}\,,
\eeq
where we have stressed the agreement with the derivation in the two-field language, and the equivalence \eqref{cs-cs}-\eqref{P(X)-A}. In Ref.~\cite{Tolley:2009fg}, the expression \eqref{cs-gelaton} of the sound speed was derived from the two-full picture, and its equivalence with the $k$-inflationary result was shown for the special case of an hyperbolic field space with $b(\chi)=g \chi/\Mp$. Here, the equivalence is shown in generic models. More importantly, Eq.~\eqref{A-P(X)} provides a non-trivial consistency check of the more general expression \eqref{A} of $\A$ in terms of multifield quantities.


\section{Conclusions} \label{sec:conclusion}

In this work we have presented the complete cubic action for fluctuations in a general two-field non-linear sigma model of inflation, written in comoving gauge in terms of the curvature perturbation $\zeta$ and the entropic mode $\F$, and expressed in a way that makes manifest the size of the contribution of each operator to the three-point correlation functions. The outcome is therefore essentially the generalization of Maldacena's result to two scalar fields with non-canonical kinetic terms. Our form of the action is interesting as it highlights the role of the various parameters that are unique to the multi-field context, such as the bending $\eta_\perp$ and the curvature of the internal field space, which have been recognized to be crucial in several novel inflationary scenarios. Along the way, we also identified and clarified some potential issues related to the contributions to correlation functions of boundary terms in the action, and we explained in particular that the usual approach of performing a field redefinition to remove such terms can be ambiguous. Given their generality, we expect our results to open the door to a wide range of applications. In particular, we believe that the action written in the form we have derived is very well suited to analytical approximations, for instance under a slow-varying approximation that is manifest in our result, or to study transient violations of it in models with features. It would also be interesting to use our form of the action in terms of the curvature perturbation to complement existing numerical methods to compute the primordial bispectrum.

A first important application that we have studied in detail is the effective single-field description that is valid when the entropic mode is heavy and may be integrated out. We showed how the EFT for the curvature mode can be very directly derived, at leading order in the derivative expansion, from our general two-field action. The resulting effective action includes all contributions from slow-roll parameters as well as all the other background coefficients of the UV completion. In particular, we have derived a general expression for the cubic Wilson coefficient $A$ that includes contributions that had not been taken into account in full generality so far in the literature, namely the third derivative of the potential and terms proportional to the field space curvature. As a non-trivial check of these results, we showed explicitly how the single-field EFT can be recast as a model of k-inflation, which we further verified by comparing with a general class of two-field models for which the EFT can be derived in terms of the full inflaton field (as opposed to only its fluctuation). Our calculations in this context can be extended in various directions, some of which we plan to tackle in future work. For instance, it would be interesting to work out the effective action to higher orders in the derivative expansion. This would provide an EFT with a wider range of validity, which may be of potential use in the analysis of multi-field theories featuring non-trivial dynamics of perturbations around the time of horizon crossing. A higher derivative EFT could also be useful to better understand the relation between more general single-field models, such as galileon inflation \cite{Burrage:2010cu}, and multi-field UV completions. Another case worth exploring is when the entropic fluctuation is light but may nevertheless be integrated out due to a large bending, giving rise to a different type of EFT characterized by a modified dispersion relation \cite{Cremonini:2010ua,Baumann:2011su,Gwyn:2012mw,Gwyn:2014doa}. Our general two-field cubic action would then provide a straightforward way to compute the relevant Wilson coefficients in such a set-up.

\begin{acknowledgments}

We are grateful to Xingang Chen, Guillaume Faye, Jacopo Fumagalli, David Mulryne, Gonzalo Palma, John Ronayne, Evangelos Sfakianakis and Dong-Gang Wang for useful discussions, as well as all the participants of the workshop Inflation \& Geometry at IAP, Paris, for extensive discussions and feedback. S.GS is supported by the European Research Council under the European Community's Seventh Framework Programme (FP7/2007-2013 Grant Agreement no.\ 307934, NIRG project). L.P and S.RP are supported by the European Research Council (ERC) under the European Union's Horizon 2020 research and innovation programme (grant agreement No 758792, GEODESI project). 

\end{acknowledgments}


\appendix

\section{Two-field cubic action} \label{sec:app-cubic-simple}


\subsection{First building blocks}
\label{expansion}

We first calculate separately the three contributions to the action \eqref{action-ADM}: the GR action in the first line, as well as for the scalar sector in the second line, the kinetic term and the potential. Writing $S=\int {\rm d}t {\rm d}^3 x {\cal L}$, and after substituting the linear solution \eqref{solution-constraints}, \textit{i.e.} $\alpha=\frac{\dot{\zeta}}{H}, \theta=-\frac{\zeta}{H}+\chi$ where $\frac{1}{a^2}\,\partial^2\chi=\epsilon\dot{\zeta}+\frac{\dot{\sigma}\eta_{\perp}}{\Mp^2}\,\F$, they read:
\beq\bal
\Lag_{\rm GR}^{(3)}=\frac{\Mp^2}{2}\,a^3\bigg[-9H^2\epsilon\zeta^3-\frac{2\epsilon}{a^2}\,\zeta (\partial \zeta)^2+\frac{1}{a^4}\left(3\zeta-\frac{\dot{\zeta}}{H}\right)\left(\partial_i\partial_j\theta\partial_i\partial_j\theta-(\partial^2\theta)^2\right)-\frac{4}{a^4}\,\partial_i\zeta\partial_i\theta\partial^2\theta\bigg]\,,
\eal
\label{LGR3}
\eeq
\beq\bal
\Lag_{\rm pot}^{(3)}&=-a^3\bigg[\frac{V_{;sss}}{6}\,\F^3+\frac{V_{;ss}}{2}\left(\frac{\dot{\zeta}}{H}+3\zeta\right)\F^2+V_{,s}\left(\frac{3\zeta\dot{\zeta}}{H}+\frac{9}{2}\,\zeta^2\right)\F+\frac{9}{2}\,V\left(\frac{\zeta^2\dot{\zeta}}{H}+\zeta^3\right)\bigg]\\
\eal
\label{Lpot3}
\eeq
\beq\bal
\Lag_{\rm kin}^{(3)}&=a^3\bigg\{\left[\frac{1}{3}\,\dot{\sigma}H\eta_{\perp}R_{\rm fs}-\frac{\dot{\sigma}^2}{12}\,R_{{\rm fs},s}\right]\F^3-\frac{1}{2}\left(\frac{\dot{\zeta}}{H}-3\zeta\right)\left[\dot{\F}^2+\left(H^2\eta_{\perp}^2-\frac{\dot{\sigma}^2}{2}\,R_{\rm fs}\right)\F^2\right]\\
&\quad -\frac{1}{2a^2}\left(\frac{\dot{\zeta}}{H}+\zeta\right)(\partial \F)^2-\frac{1}{a^2}\,\dot{\F}\partial \F\partial \theta-\dot{\sigma}H\eta_{\perp}\left(\frac{\dot{\zeta}^2}{H^2}-\frac{3\zeta\dot{\zeta}}{H}+\frac{9}{2}\,\zeta^2\right)\F\\
&\quad -\frac{\dot{\sigma}^2}{2}\left(\frac{\dot{\zeta}^3}{H^3}-\frac{3\zeta\dot{\zeta}^2}{H^2}+\frac{9\zeta^2\dot{\zeta}}{2H}-\frac{9}{2}\,\zeta^3\right)\bigg\}\,,
\eal
\label{Lkin3}
\eeq
supplemented by the boundary term \eqref{D0}, and where for the kinetic term it may be useful to first expand the tensor ${\cal G}_{\mu\nu}\equiv G_{IJ}(\phi)\partial_{\mu}\phi^I\partial_{\nu}\phi^J$ to cubic order in perturbations:
\beq\bal
{\cal G}_{00}&=\dot{\sigma}^2-2\dot{\sigma}H\eta_{\perp}\F+\dot{\F}^2+H^2\eta_{\perp}^2\F^2-\frac{\dot{\sigma}^2}{2}\,R_{\rm fs}\F^2+\frac{2}{3}\,\dot{\sigma}H\eta_{\perp}R_{\rm fs}\F^3-\frac{\dot{\sigma}^2}{6}\,R_{{\rm fs},s}\F^3\,,\\
{\cal G}_{0i}&=\dot{\F}\partial_i\F\,,\qquad {\cal G}_{ij}=\partial_i\F\partial_j\F\,.
\eal\eeq
Summing the three contributions \eqref{LGR3}-\eqref{Lkin3} and using the background equations, one arrives at the simple form \eqref{L3-naive} of the cubic action. As explained in the main text, many manipulations should be performed in order to render explicit the true size of the cubic couplings. These are presented in the next subsection.


\subsection{Manipulations of the cubic action}
\label{manipulations}

To structure the computation, we split the initial form \eqref{L3-naive} of the cubic action into
\beq
\mathcal{L}^{(3)}=\mathcal{L}_{{\rm ini}}^{(3)}(\zeta,\theta) + \mathcal{L}_{{\rm ini}}^{(3)}(\zeta,\mathcal{F}) + \mathcal{D}_0
\label{L3-split}
\eeq
with $\mathcal{L}_{{\rm ini}}^{(3)}(\zeta,\theta)$ the part coming (mostly) from the GR action, whose dependence on $\F$ only comes through solving the constraint equations and hence the $\F$-dependence of $\theta$, \textit{i.e.}
\beq\bal
\mathcal{L}_{{\rm ini}}^{(3)}(\zeta,\theta)&=a^3 \Mp^2\bigg[\epsilon\left(3\zeta-\frac{\dot{\zeta}}{H}\right)\dot{\zeta}^2-\frac{\epsilon}{a^2}\,\zeta (\partial\zeta)^2+\frac{1}{2a^4}\left(3\zeta-\frac{\dot{\zeta}}{H}\right)\left(\partial_i\partial_j\theta\partial_i\partial_j\theta-(\partial^2\theta)^2\right)\\
&\quad \quad \quad \quad -\frac{2}{a^4}\,\partial_i\zeta\partial_i\theta\partial^2\theta\bigg]\,,
\eal
\label{L3-ini-GR}
\eeq
and $\mathcal{L}_{{\rm ini}}^{(3)}(\zeta,\mathcal{F})$ that has no single-field counterpart and originates (mostly) from the scalar kinetic and potential terms of the action, \textit{i.e.}
\beq\bal
\mathcal{L}_{{\rm ini}}^{(3)}(\zeta,\F)&=a^3\bigg\{ -\frac{1}{a^2}\,\dot{\F}\partial \F\partial \theta+\frac{1}{2}\left(3\zeta-\frac{\dot{\zeta}}{H}\right)\dot{\F}^2-\frac{1}{2a^2}\left(\zeta+\frac{\dot{\zeta}}{H}\right)(\partial \F)^2\\
&\quad+ \dot{\sigma}\eta_{\perp}\left(6\zeta-\frac{\dot{\zeta}}{H}\right)\dot{\zeta}\F-\frac{1}{2H}\Big(m_s^2+2H^2\eta_{\perp}^2-2\epsilon \Mp^2H^2R_{\rm fs}\Big)\dot{\zeta}\F^2-\frac{3}{2}\,m_s^2\zeta\F^2\\
&\quad -\frac{1}{6}\left(V_{;sss}-2\dot{\sigma}H\eta_{\perp}R_{\rm fs}+\epsilon \Mp^2H^2R_{{\rm fs},s}\right)\F^3\bigg\}\,.
\eal
\label{Lini-F}
\eeq


\subsubsection{Manipulations of $\mathcal{L}_{{\rm ini}}^{(3)}(\zeta,\theta)$}

Similarly to Ref.~\cite{Collins:2011mz}, we consider separately terms in \eqref{L3-ini-GR} with different powers of $a$. Our computation is different in practice, as we treat on equal footing the two contributions to $\theta=-\frac{\zeta}{H}+\chi$, and, as much as possible, do not split $\chi$ into its two contributions in \eqref{def-chi}. This simplifies the computation even in the single-field framework, and it enables one to extend it more easily to the two-field situation. We thus divide $\mathcal{L}_{{\rm ini}}^{(3)}(\zeta,\theta)$ into four groups of terms:
\begin{align}
\mathcal{L}_{{\rm ini}}^{{\rm (3),\, I}}(\zeta)&=a^3 \Mp^2\, \epsilon  \left(3\zeta-\frac{\dot{\zeta}}{H}\right)\dot{\zeta
}^2  \\
\mathcal{L}_{{\rm ini}}^{{\rm(3),\,II}}(\zeta)&= - a \Mp^2\, \epsilon \, \zeta \left(\partial\zeta\right)^2 \\
\mathcal{L}_{{\rm ini}}^{{\rm(3),\,III.B.}}(\zeta,\theta)&= \frac{\Mp^2}{2a}   \left(3\zeta-\frac{\dot{\zeta}}{H} \right) \left(\partial_{ij}\theta\partial_{ij}\theta - (\partial^2\theta)^2\right)\\
\mathcal{L}_{{\rm ini}}^{{\rm(3),\,III.A.}}(\zeta,\theta)&= -\frac{2}{a}  \Mp^2\, \partial^2 \theta \left(\partial\theta\right) \left(\partial \zeta\right)\,, \label{fourth-group}
\end{align}

\noindent {\bf First group}.--- Integrating by parts, one simply writes:
\begin{align}
\mathcal{L}_{{\rm ini}}^{{\rm (3),\, I}}(\zeta)&=a^3\Mp^2 \epsilon (\epsilon-\eta)\dot{\zeta}^2\zeta+\frac{2}{H}\zeta \dot{\zeta} \frac{d}{dt}\left(a^3 \Mp^2\epsilon\dot{\zeta}\right) +\mathcal{D}_{\rm I} \\
\text{with } \mathcal{D}_{\rm I}&=-\frac{d}{dt}\left[a^3 \Mp^2 \frac{\epsilon}{H}\dot{\zeta}^2 \zeta \right]
\end{align}

\noindent {\bf Second group}.--- No manipulation is needed on the second group. However we will show later that it combines with other terms to give a term proportional to the equation of motion for $\zeta$.

\noindent {\bf Third and fourth groups}.--- More manipulations are needed here. Using $\theta=-\frac{\zeta}{H}+\chi$, the third group reads:

\beq\bal
\mathcal{L}_{{\rm ini}}^{{\rm (3),\,III.B.}}(\zeta,\theta)&=\frac{\Mp^2}{2a} \left(3\zeta-\frac{\dot{\zeta}}{H}\right) \bigg[ \frac{1}{H^2} \left(\zeta_{,ij}\zeta_{,ij}-\left(\partial^2 \zeta\right)^2 \right) - \frac{2}{H}\left(\zeta_{,ij}\chi_{,ij}-\partial^2\zeta \partial^2 \chi \right) \\
&\qquad \qquad \qquad \qquad \quad+ \left(\chi_{,ij}\chi_{,ij}-\left(\partial^2\chi\right)^2 \right) \bigg]
\label{third-group-developed}
\eal
\eeq
Performing temporal integrations by parts, and using \eqref{zeta-eom-chi}, one can derive the useful identities, for any background functions $f_i(t)$:
\begin{align}
f_1(t)\dot{\zeta}\left(\zeta_{,ij}\zeta_{,ij}-\left(\partial^2 \zeta\right)^2 \right)&=\frac{1}{3}\frac{d}{dt}\left[f_1\dot{\zeta}\left(\zeta_{,ij}\zeta_{,ij}-\left(\partial^2 \zeta\right)^2 \right)\right]- \frac{1}{3} \dot{f_1} \zeta \left(\zeta_{,ij}\zeta_{,ij}-\left(\partial^2 \zeta\right)^2 \right) \\
f_2(t)\dot{\zeta}\left(\zeta_{,ij}\chi_{,ij}-\partial^2 \zeta\partial^2\chi \right)&=\frac{1}{2}\frac{d}{dt}\left[f_2\dot{\zeta}\left(\zeta_{,ij}\chi_{,ij}-\partial^2 \zeta\partial^2 \chi \right)\right] - \frac{\epsilon}{2} f_2\zeta\left(\zeta_{,ij}\zeta_{,ij}-\left(\partial^2 \zeta\right)^2 \right)  \nonumber \\
&\hspace{-8em}- \frac{1}{2} \left( \dot{f}_2 - f_2H\right) \zeta \left(\zeta_{,ij}\chi_{,ij}-\partial^2 \zeta\partial^2 \chi \right)  - \frac{a^2 f_2}{4 \Mp^2}\zeta\left(\zeta_{,ij}\partial_i\partial_j\partial^{-2}\cdot- \partial^2 \zeta \right) \Ez\\
f_3(t)\dot{\zeta}\left(\chi_{,ij}\chi_{,ij}-\left(\partial^2 \chi\right)^2 \right)&=\frac{d}{dt}\left[f_3\dot{\zeta}\left(\chi_{,ij}\chi_{,ij}-\left(\partial^2 \chi\right)^2 \right)\right]
- 2 \epsilon f_3 \left( \zeta_{,ij}\chi_{,ij}-\partial^2 \zeta\partial^2 \chi  \right) \nonumber \\
&\hspace{-8em} - \left(\dot{f}_3 -2f_3H\right) \zeta \left(\chi_{,ij}\chi_{,ij}-\left(\partial^2 \chi\right)^2 \right)   - \frac{a^2 f_3}{\Mp^2}\zeta\left(\chi_{,ij}\partial_i\partial_j\partial^{-2}\cdot- \partial^2 \chi \right) \Ez\,.
\end{align}
Applying these relations to \eqref{third-group-developed}, with $f_1(t)=-\frac{\Mp^2}{2aH^3}$ , $f_2(t)=\frac{\Mp^2}{aH^2}$ and $f_3(t)=-\frac{\Mp^2}{2aH}$, 
and using the identity
\beq \label{eq:eft spatial ibp identity 2}
\zeta \left( \partial_i\partial_jg \partial_i\partial_j h - \partial^2g\partial^2 h \right)=\frac{1}{2}\,\partial^2\zeta\partial_i g \partial_i h+\frac{1}{2}\left(\partial^2 g\partial_i\zeta\partial_i h +\partial^2 h\partial_i\zeta\partial_i g\right)\,
\eeq
valid for any functions $g$ and $h$, and where total spatial derivative are discarded, to simplify bulk terms, one obtains:
\begin{align}
\mathcal{L}_{{\rm ini}}^{{\rm (3),\, III.B.}}(\zeta,\theta)=&\frac{2 \Mp^2}{aH^2} \left[\left(\partial\zeta\right)^2\partial^2 \zeta \right] - \frac{\Mp^2}{aH} \left[\left(\partial\zeta\right)^2 \partial^2 \chi + 2 \left(\partial\zeta\right)\left(\partial\chi\right) \partial^2 \zeta \right]\\
&+ \frac{\epsilon \Mp^2}{4a}\left[\left(\partial\chi\right)^2 \partial^2 \zeta+2\left(\partial\zeta\right)\left(\partial\chi\right) \partial^2 \chi \right]  + \mathcal{D}_{{\rm III.B.}}+\eom_{{\rm III.B.}}
\end{align}
with
\beq\bal
\mathcal{D}_{{\rm III.B.}}&=- \Mp^2\frac{d}{dt} \bigg[  \frac{1}{6aH^3} \zeta \left( \zeta_{,ij}\zeta_{,ij}-(\partial^2\zeta)^2 \right) - \frac{1}{2aH^2} \zeta\left(\zeta_{,ij}\chi_{,ij}-\partial^2\zeta \partial^2\chi\right) \\
&\qquad \qquad  \quad+ \frac{1}{2aH} \zeta \left(\chi_{,ij}\chi_{,ij}- \left(\partial^2\chi \right)^2 \right)   \bigg]
\eal
\eeq
and
\begin{align}
\eom_{{\rm III.B.}}&=-\frac{a \zeta}{4 H^2} \left[  \left(\zeta_{,ij}\partial_i\partial_j\partial^{-2} - \partial^2 \zeta \right) -2H \left(\chi_{,ij}\partial_i\partial_j\partial^{-2} - \partial^2 \chi \right) \right]  \Ez\,.
\end{align}
Developing the expression of the fourth group \eqref{fourth-group} in terms of $\zeta$ and $\chi$, it reads
\begin{equation}
\mathcal{L}_{{\rm ini}}^{{\rm (3),\, III.A.}}(\zeta,\theta)=-\frac{2 \Mp^2}{aH^2} \left[\left(\partial \zeta\right)^2 \partial^2 \zeta \right] + \frac{2  \Mp^2}{aH}\left[ \left(\partial\zeta\right)^2 \partial^2 \chi + \left(\partial\zeta\right)\left(\partial\chi\right)\partial^2 \zeta \right] - \frac{2  \Mp^2}{a} \left[ \left(\partial\zeta\right)\left(\partial\chi\right) \partial^2 \chi \right]\,,
\end{equation}
from which one deduces the rather compact form
\begin{align}
\mathcal{L}_{{\rm ini}}^{{\rm (3),\, III.A.}}(\zeta,\theta)+\mathcal{L}_{ini}^{{\rm (3),\,III.B.}}(\zeta,\theta)&= \frac{\Mp^2}{aH} \left(\partial\zeta\right)^2 \partial^2 \chi + \frac{\Mp^2}{a}\left[ \left(\frac{\epsilon}{2}-2\right) \left(\partial\zeta\right)\left(\partial\chi\right) \partial^2 \chi + \frac{\epsilon}{4} \left(\partial\chi\right)^2 \partial^2 \zeta \right] \nonumber \\
&+ \mathcal{D}_{{\rm III.B.}}+\eom_{{\rm III.B.}}\,,
\label{crucial-step}
\end{align}
where the expression of $\chi$ \eqref{def-chi} may now be used to express all quantities explicitly in terms of $\zeta$ and $\F$.

\noindent {\bf Addition and manipulation of the four groups}.--- Adding the four groups all together and using
\beq\bal
\frac{2}{H}\zeta \dot{\zeta} \frac{d}{dt}\left(a^3\Mp^2\epsilon\dot{\zeta}\right) +  a\epsilon \Mp^2\left( \frac{\dot{\zeta}}{H}- \zeta \right) \left(\partial\zeta\right)^2& =
 a \epsilon \Mp^2(\epsilon+\eta) \zeta (\partial \zeta)^2
- \frac{2}{H} \dot{\zeta}\zeta  \frac{d}{dt}\left(a^3 \dot{\sigma}\eta_\perp\F\right) \\
&- \frac{d}{dt}\left[\frac{a\epsilon \Mp^2}{H} \zeta \left(\partial\zeta\right)^2 \right] +  \frac{a^3}{H} \dot{\zeta}\zeta \Ez\,,
\eal
\eeq
one eventually obtains
\beq \bal
\mathcal{L}_{{\rm ini}}^{(3)}(\zeta,\theta)&= a^3 \Mp^2\left[ \epsilon (\epsilon-\eta) \dot{\zeta}^2\zeta+ \epsilon (\epsilon+\eta)\zeta \frac{\left(\partial \zeta \right)^2}{a^2}  +
\left(\frac{\epsilon}{2}-2\right)\frac{1}{a^4} \left(\partial\zeta\right)\left(\partial\chi\right) \partial^2 \chi + \frac{\epsilon}{4 a^4} \partial^2 \zeta ( \partial \chi)^2 \right]\\
&+ a\eta_\perp \frac{\dot{\sigma}}{H} \mathcal{F} (\partial\zeta)^2 - \frac{2}{H} \dot{\zeta} \zeta \frac{d}{dt} \left(a^3 \dot{\sigma}\eta_\perp \mathcal{F} \right)+\mathcal{D}_1+\eom_1
\label{Lini-result}
\eal
\eeq
with
\bea
\mathcal{D}_{1}&=&- \Mp^2\frac{d}{dt} \bigg[ a^3 \frac{\epsilon}{H} \zeta \dot{\zeta}^2+ a \frac{\epsilon}{H}\zeta (\partial\zeta)^2 + \frac{1}{6aH^3} \zeta \left( \zeta_{,ij}\zeta_{,ij}-(\partial^2\zeta)^2 \right) - \frac{1}{2aH^2} \zeta\left(\zeta_{,ij}\chi_{,ij}-\partial^2\zeta \partial^2\chi\right) \nonumber \\
&&+ \frac{1}{2aH} \zeta \left(\chi_{,ij}\chi_{,ij}- \left(\partial^2\chi \right)^2 \right)   \bigg] \\
\eom_1&=&\left\{ \frac{a^3}{H}\dot{\zeta} \zeta -\frac{a \zeta}{4 H^2} \left[  \left(\zeta_{,ij}\partial_i\partial_j\partial^{-2} - \partial^2 \zeta \right) -2H \left(\chi_{,ij}\partial_i\partial_j\partial^{-2} - \partial^2 \chi \right) \right]\right\} \Ez
\eea
The form \eqref{Lini-result} of the cubic action (supplemented with ${\cal D}_0$ \eqref{D0}) reproduces the single-field result when $\F$ is absent. We chose to organize the vertices and the boundary terms in the same manner as in Ref.~\cite{Burrage:2011hd} (see Eqs.~3.2 and 3.10 there).
As explained in detail in the main text, this form of the action is interesting because, in the single-field case where $\zeta$ becomes constant on super-Hubble scales, the displayed boundary terms do not contribute to the bispectrum, and the number of operators in the bulk action is minimized.
More importantly, all manipulations have been made so that the dynamically relevant terms in \eqref{Lini-result} are manifestly of order ${\cal O}(\epsilon,\eta)^2$ or higher. In the more general two-field case of interest, similar manipulations should be performed for the terms involving $\F$, to which we now turn.


\subsubsection{Manipulating the interactions involving entropic perturbations and total cubic action}
\label{manipulations-F}

The cubic interactions involving $\F$ in \eqref{Lini-F} and \eqref{Lini-result} and that do not manifestly display the right amplitude of the interactions all appear in the first line of \eqref{Lini-F}. These terms appear without $\epsilon$ factors, nor background parameters related to the entropic sector like $\eta_\perp, m_s^2, R_{\rm fs}$ and similar). The manipulations that remedy this are
\begin{align}
\frac{a}{2}\left[\frac{2}{H}\dot{\mathcal{F}}\partial \mathcal{F} \partial \zeta- (\partial\mathcal{F})^2 \left(\zeta+\frac{\dot{\zeta}}{H} \right) \right]&=\frac{a}{2}\left[\epsilon\zeta\left(\partial\F\right)^2- \frac{2}{H}\zeta\dot{\F} \partial^2\F \right]-\frac{d}{dt}\left[\frac{a}{2H}\left(\partial\F\right)^2 \zeta\right]
\label{simplification1}
\end{align}
and
\begin{align}
\frac{a^3}{2}\left(3\zeta-\frac{\dot{\zeta}}{H}\right)\dot{\F}^2&=\frac{a^3}{2}\epsilon\zeta\dot{\F}^2+\frac{a}{H} \zeta \dot{\F} \partial^2\F -\frac{a^3m_s^2}{H}\dot{\F}\F\zeta+\frac{2a^3\dot{\sigma}\eta_\perp}{H}\dot{\zeta}\zeta
\dot{\F}-\frac{d}{dt}\left(\frac{a^3}{2H}\zeta\dot{\F}^2\right)+\frac{a^3}{H} \zeta \dot{\F} {\cal E}_{\F}
\label{simplification2}
\end{align}
where the terms in $\zeta \dot{\F} \partial^2\F$ in the right hand sides cancel in the total action. To put the action in the best form that we give in the main next, this is not enough though, notably because the generated terms proportional to $m_s^2\dot{\F}\F\zeta$ and $\eta_\perp \dot{\zeta}\zeta \dot{\F}$ contribute to the cubic action at leading order in situations where $\F$ is integrated out (see section \ref{regime-of-validity}), and we prefer that terms involving $\dot{\F}$ are readily negligible, to make the derivation of the single-field EFT more transparent. For this purpose, we note that, integrating by parts the term in $m_s^2\dot{\F}\F\zeta$, several cancellations arise amongst different terms to obtain
\beq \bal
&a^3 \dot{\sigma}\eta_{\perp}\left(6\zeta-\frac{\dot{\zeta}}{H}\right)\dot{\zeta}\F - \frac{d}{dt} \left(a^3 \dot{\sigma}\eta_\perp \mathcal{F} \right)\left(\frac{2}{H} \dot{\zeta} \zeta \right) -\frac{a^3m_s^2}{H}\dot{\F}\F\zeta+\frac{2a^3\dot{\sigma}\eta_\perp}{H}\dot{\zeta}\zeta-\frac{3}{2}\,a^3 m_s^2\zeta\F^2 \\
&=a^3 \dot\sigma \eta_\perp(2 \epsilon-\eta-2 \lambda_\perp) \dot\zeta \zeta \F+\frac{a^3 m_s^2}{2 H} \left(\F-\frac{2 \dot\sigma \eta_\perp}{m_s^2} \dot\zeta \right) \dot\zeta \F+\frac12 a^3 m_s^2(\epsilon+\mu_s) \zeta \F^2-\frac12 \frac{d}{dt}\left(a^3 \frac{m_s^2}{H} \zeta \F^2 \right)
\label{simplification3}
\eal\eeq
 where no derivative of $\F$ appears in the right hand side except for the boundary term, and we defined $\lambda_\perp=\dot{\eta}_\perp/(H \eta_\perp)$ and $\mu_s=\Tdot{(m_s^2)}/(H m_s^2)$. Now summing all the contributions  to \eqref{L3-split}, \textit{i.e.} \eqref{D0}, \eqref{Lini-result}, \eqref{Lini-F} and using Eqs.~\eqref{simplification1}-\eqref{simplification3}, one finds:
 \beq\bal \label{eq:cubic action best form1}
\Lag^{(3)}&= \Mp^2 \,a^3\left[ \epsilon (\epsilon-\eta) \dot{\zeta}^2\zeta+ \epsilon (\epsilon+\eta)\zeta \frac{\left(\partial \zeta \right)^2}{a^2}  +
\left(\frac{\epsilon}{2}-2\right)\frac{1}{a^4} \left(\partial\zeta\right)\left(\partial\chi\right) \partial^2 \chi + \frac{\epsilon}{4 a^4} \partial^2 \zeta ( \partial \chi)^2 \right] \\
&+a^3\bigg[ \frac12 m_s^2 (\epsilon+\mu_s)\zeta \F^2+\left(2\epsilon-\eta-2\lambda_{\perp}\right)\dot\sigma \eta_\perp \zeta  \dot\zeta  \F
+\frac{\dot{\sigma}\eta_{\perp}}{a^2H}\,\F(\partial\zeta)^2\\
& -\frac{\dot{\sigma}\eta_{\perp}}{H}\,\dot{\zeta}^2\F -\frac{1}{H}\left(H^2\eta_{\perp}^2-\epsilon \Mp^2H^2R_{\rm fs}\right)\dot{\zeta}\F^2-\frac{1}{6}\left(V_{;sss}-2\dot{\sigma}H\eta_{\perp}R_{\rm fs}+\epsilon \Mp^2H^2R_{{\rm fs},s}\right)\F^3\\
&+\frac12\epsilon\zeta \left(\dot{\F}^2+\frac{(\partial\F)^2}{a^2}\right)- \frac{1}{a^2}\dot{\F}\partial \F\partial  \chi \bigg] +{\cal D}+{\cal E}
\eal
\eeq
with
\beq\bal  \label{D}
{\cal D}&= \frac{d}{dt}\bigg\{-\frac{a}{2 H}\zeta(\partial\F)^2+\frac{a \Mp^2}{H}(1-\epsilon)\zeta(\partial\zeta)^2+a^3\bigg[-9H \Mp^2\zeta^3-\frac{\epsilon \Mp^2}{H}\,\dot{\zeta}^2\zeta -\frac{1}{2H}\,\zeta\dot{\F}^2-\frac{m_s^2}{2H}\,\zeta\F^2\bigg] \\
&\quad-\frac{\Mp^2 \zeta}{6 aH^3} \left(\left(\zeta_{,ij}\zeta_{,ij}-(\partial^2\zeta)^2\right) -3 H\left(\zeta_{,ij} \chi_{,ij}-\partial^2\zeta\partial^2\chi\right)+3 H^2\left(\chi_{,ij} \chi_{,ij}-(\partial^2\chi)^2\right) \right)\bigg\}\,,  \\
{\cal E}&=\frac{a^3}{H}\,{\cal E}_{\zeta}\bigg[\dot{\zeta}\zeta-\frac{1}{4a^2H}\left( (\partial \zeta)^2-\partial^{-2}\partial_i\partial_j(\partial_i\zeta\partial_j\zeta) -2H\left(\partial \zeta\partial \chi-\partial^{-2}\partial_i\partial_j(\partial_i\zeta\partial_j \chi)\right) \right)\bigg]+\frac{a^3}{H}\,{\cal E}_{\F}\zeta\dot{\F}\,.
\eal
 \eeq
The compact expression of the cubic action \eqref{eq:cubic action best form1} with the boundary term \eqref{D} is the main result of this paper. Its usefulness and its consequences are discussed in the main text, where we simply slightly changed the appearance of the boundary term in \eqref{D-main-text} to make appear conjugate momenta, as this simplifies subsequent calculations.


\section{Boundary terms and field redefinitions} \label{sec:field-redefinitions}

In this Appendix we briefly review the method of performing non-linear field redefinitions to compute the contributions to cosmological correlators of boundary terms in the action (see \textit{e.g.}~\cite{Burrage:2011hd}; see also \cite{Goon:2018fyu} for a related discussion in the context of the wave function approach). In short, the idea of the method is to redefine field variables so that the relevant boundary terms disappear, and then to simply work out the relation between correlators of the new and old variables. Although this procedure has become standard in the single-field context, we would like to highlight some potential ambiguities that may lead to incorrect results, especially in the more complicated multi-field set-up. It is mainly for this reason that we chose the more direct method of Sec.\ \ref{sec:boundary terms} to compute the contributions of boundary terms.

Consider a generic quadratic field redefinition
\beq \label{eq:generic field redef}
\zeta=\tilde{\zeta}+f[\tilde{\zeta},\tilde{\F}]\,,\qquad \F=\tilde{\F}+g[\tilde{\zeta},\tilde{\F}]\,,
\eeq
where, by assumption, the functionals $f$ and $g$ are quadratic in the fields. It then follows that the quadratic part \eqref{L2} of the action reads, in terms of the new variables:
\beq\bal \label{eq:field redef general}
S^{(2)}\left[\zeta,\F\right]&=S^{(2)}[\tilde{\zeta},\tilde{\F}]-\int d^4x\,a^3\big[f{\cal E}_{\zeta}+g{\cal E}_{\F}\big] +\int d^4x\,\partial_t\left[2\Mp^2a f \partial^2 \tilde \chi+a^3g\dot{\tilde{\F}}\right]+\cdots
\eal\eeq
where the ellipses stand for quartic terms. Recalling that terms proportional to the linear equations of motion ${\cal E}_{\zeta}$ and ${\cal E}_{\F}$ are irrelevant, one can see that the only impact of field redefinitions of $\zeta$ and $\F$ is to introduce boundary terms, proportional to $\partial^2 \chi$ and $\dot{\F}$ respectively.\footnote{The fact that the generated terms, in $\partial^2 \chi$ and $\dot{\F}$, are proportional to the (linear) conjugate momenta of $\zeta$ and $\F$ is in agreement with the discussion below Eq.~\eqref{contribution-boundary-term}, where the commutator form readily indicates that only boundary terms proportional to conjugate momenta contribute to correlation functions of fields.}
We precisely organised the boundary term \eqref{D-main-text} in that way, with the first line containing only operators with no time derivatives (hence which do not affect correlation functions of fields), and the last two lines containing the terms that can be removed via field redefinitions. Indeed, choosing the functions $f$ and $g$ as (note that at this order, writing them in terms of the original $(\zeta,\F)$ is irrelevant)
\beq\bal
f[\zeta,\F]&= \frac{ \zeta \dot\zeta}{H}-\frac{1}{8 a^2 H^2} \left(  (\partial \zeta)^2-2\partial_i \partial^{-2}(\partial^i \zeta \partial^2 \zeta) -2H (\partial \zeta \partial \chi-\frac12 \partial_i \partial^{-2}(\partial^i \chi \partial^2 \zeta))+\frac{4H}{\epsilon} \zeta \partial^2 \chi \right)\,,\\
g[\zeta,\F]&=\frac{1}{2H}\,\zeta\dot{\F}\,,
\label{f-g}
\eal\eeq
the action up to cubic order in terms of the variables $(\tilde \zeta,\tilde \F)$ contain only the bulk interactions in \eqref{eq:cubic action best form1-main-text}, whose contributions to correlation functions can be calculated using standard methods, with analytical approximations or numerically. Eventually, we just have to take into account the difference between three-point functions of the original and redefined variables \eqref{eq:generic field redef}, which is straightforward to do using Wick's theorem.

The tricky aspect of this method is the fact that the above field redefinition is not unique, simply because the structure of the boundary term \eqref{D-main-text} can be changed by doing spatial integrations by parts. To illustrate this, consider the terms of the form $\zeta p_\zeta^2$ in \eqref{D-main-text}, which in Fourier space will read ${\cal L}\supset h(k_1,k_2,k_3)\zeta(k_1)p_{\zeta}(k_2)p_{\zeta}(k_3)$ for some function $h$ of the wave vectors. However, in order to read off the field redefinition that is supposed to handle this term, one needs to remove all spatial derivatives of one of the field momenta (say $p_\zeta(k_3)$) so that, when written in this form, the function $h$ becomes independent of $k_3$:
\beq \label{eq:h1 bdy term}
h(k_1,k_2,k_3)\zeta(k_1)p_{\zeta}(k_2)p_{\zeta}(k_3)=h_1(k_1,k_2)\zeta(k_1)p_{\zeta}(k_2)p_{\zeta}(k_3)+{\rm t.d.}\,.
\eeq
A field redefinition that removes this expression is given by taking $f=h_1(k_1,k_2)\zeta(k_1)p_{\zeta}(k_2)$ (with a convolution over the wave vectors being implicit when working in Fourier space). However, as we have stressed, this is not unique because \eqref{eq:h1 bdy term} is not unique: by doing an integration by parts we have
\beq
h_1(k_1,k_2)\zeta(k_1)p_{\zeta}(k_2)p_{\zeta}(k_3)=h_2(k_1,k_2)\zeta(k_1)p_{\zeta}(k_2)p_{\zeta}(k_3)+{\rm t.d.}\,,
\eeq
with a new function $h_2$ and hence a different function $f$ in \eqref{eq:generic field redef}. Since it is precisely the function $f$ that determines the relation between the correlators of $\tilde\zeta$ and the correlators of $\zeta$, we see that this procedure can lead to ambiguous results.


\bibliography{Revisiting-NG-JHEP2.bbl}

\end{document}